\documentclass[]{aa}  
\usepackage{graphicx}
\usepackage{txfonts}

\usepackage[colorlinks=true,linkcolor=blue, citecolor=blue]{hyperref}
\usepackage{amsmath}
\usepackage{xcolor}

\newcommand{\Gaia}{{\it Gaia}}
\newcommand{\Feh}{$[\mathrm{Fe/H}]$}

\begin{document}

   \title{The Blanco DECam Bulge Survey (BDBS)}

   \subtitle{VIII. Chemo-kinematics in the southern Galactic bulge from 2.3 million red clump stars with {\Gaia} DR3 proper motions}

   \author{Tommaso Marchetti
          \inst{1},
          Meridith Joyce
          \inst{2,3},
          Christian I. Johnson
          \inst{4},
          R. Michael Rich
          \inst{5},
          William Clarkson
          \inst{6},
          Andrea Kunder
          \inst{7},
          Iulia T. Simion
          \inst{8},
          Catherine A. Pilachowski
          \inst{9}
          }

\authorrunning{T. Marchetti et al.}
\titlerunning{BDBS. VIII. Chemo-kinematics of Red Clump Stars}

   \institute{European Southern Observatory, Karl-Schwarzschild-Strasse 2, 85748 Garching bei München, Germany\\
              \email{tommaso.marchetti.astro@gmail.com}
         \and
            Konkoly Observatory, Research Centre for Astronomy and Earth Sciences, H-1121 Budapest Konkoly Th. M. \'ut 15-17., Hungary
         \and
            CSFK, MTA Centre of Excellence, Budapest, Konkoly Thege Mikl\'os \'ut 15-17., H-1121, Hungary
            \and
             Space Telescope Science Institute, 3700 San Martin Drive, Baltimore, MD 21218, USA
         \and
             Department of Physics and Astronomy, University of California Los Angeles, 430 Portola Plaza, Box 951547, Los Angeles, CA 90095-1547, USA
        \and
            Department of Natural Sciences, University of Michigan-Dearborn, 4901 Evergreen Rd., Dearborn, MI 48128, USA
        \and
            Saint Martin’s University, 5000 Abbey Way SE, Lacey, WA 98503, USA
        \and
             Shanghai Key Lab for Astrophysics, Shanghai Normal University, 100 Guilin Road, Shanghai 200234, PR China
        \and
            Indiana University Department of Astronomy, SW319, 727 E 3rd Street, Bloomington, IN 47405, USA
             }

   \date{Received XXX; accepted YYY}


  \abstract
  {The inner Galaxy is a complex environment, and the relative contributions of different formation scenarios to its observed morphology and stellar properties are still debated. The different components are expected to have different spatial, kinematic, and metallicity distributions, and a combination of photometric, spectroscopic, and astrometric large-scale surveys is needed to study the formation and evolution of the Galactic bulge.}
   {The Blanco DECam Bulge Survey (BDBS) provides near-ultraviolet to near-infrared photometry for $\sim 250$ million unique stars over $>$200 square degrees of the southern Galactic bulge. By combining BDBS photometry with the latest {\Gaia} astrometry, we aim to characterize the chemo-dynamics of red clump stars across the BDBS footprint, using an unprecedented sample size and sky coverage.}
   {Our field of regard is $|\ell| \leq 10^\circ$, $-10^\circ \leq b \leq -3^\circ$.  We construct a sample of $\sim 2.3$ million red clump giants in the bulge with photometric metallicities, BDBS photometric distances, and proper motions.  Photometric metallicities are derived from a $(u-i)_0 {\rm vs\  [Fe/H]}$ relation; astrometry including precise proper motions is from the third data release (DR3) of the ESA satellite {\Gaia}. 
   We study the kinematics of the red clump stars as a function of sky position and metallicity, by investigating proper motion rotation curves, velocity dispersions, and proper motion correlations across the southern Galactic bulge.}
   {By binning our sample into 8 metallicity bins from $-1.5$ dex $<$ {\Feh} $<$ $+1$ dex, we find that metal-poor red clump stars exhibit lower rotation amplitudes, at $\sim 29$ km s$^{-1}$ kpc$^{-1}$. The peak of the angular velocity is $\sim 39$ km s$^{-1}$ kpc$^{-1}$ for {\Feh} $\sim -0.2$ dex, exhibiting declining rotation at higher [Fe/H]. The velocity dispersion is higher for metal-poor stars, while metal-rich stars show a steeper gradient with Galactic latitude, with a maximum dispersion at low latitudes along the bulge minor axis. Only metal-rich stars ({\Feh} $\gtrsim -0.5$ dex) show clear signatures of the bar in their kinematics, while the metal-poor population exhibits isotropic motions with an axisymmetric pattern around Galactic longitude $\ell = 0$.
   }
   {This work reports the largest sample of bulge stars with distance, metallicity, and astrometry and shows clear kinematic differences with metallicity. The global kinematics over the bulge agrees with earlier studies.  However, we see striking changes with increasing metallicity and for the first time, see kinematic differences for stars with $\rm [Fe/H]>-0.5$, suggesting that the bar itself may have kinematics that depends on metallicity.}
   

   \maketitle
%

\section{Introduction}
\label{sec:intro}

The last decades of observations of the central region of our Galaxy have pictured a highly complex environment, with several distinct and interacting stellar populations coexisting along the same line of sight. Studying the formation, history, and evolution of the Galactic bulge is essential to improve our knowledge of the Milky Way, which is a benchmark for understanding the evolution of disk galaxies in the Universe \citep[e.g.][]{Bland-Hawthorn16}. The reconstruction of its evolutionary history requires precise photometric \citep[e.g.][]{Minniti10, Udalski15, Rich20}, spectroscopic \citep[e.g.][]{Rich07, Kunder12, Freeman13, Zoccali14, Majewski17}, and astrometric \citep[e.g.][]{Kuijken02, clarkson08,Clarkson18, gezari+22} measurements for statistically large samples of stars covering the whole Galactic bulge. Because of our location in the Milky Way plane, several studies have focused on the removal of the foreground disk population, whose presence contaminates studies of the Galactic bulge, using a mix of photometric and astrometric techniques \citep[e.g.,][]{Kuijken02, Zoccali03, clarkson08, clarkson11, Valenti13, Calamida14, Clarkson18, Surot19b, Queiroz+21, Marchetti+22}.

While the recent data releases of the European Space Agency satellite {\Gaia} \citep{Gaia16, Gaia18_DR2, GaiaEDR3, GaiaDR3} provide geometric parallaxes for $1.5$ billion stars in the Milky Way, large uncertainties and systematics prevent the precise localization of stars beyond a few kiloparsec from the Sun. The identification of stellar standard candles, whose distance can be derived using photometry, is, therefore, an important tool to map the three-dimensional structure of the Galaxy. In particular, Red Clump (RC) stars have long been used to study and constrain the morphology of the inner part of the Milky Way \citep[e.g.][]{Stanek94, Stanek97, Rattenbury+07, McWilliam10, Saito11, Wegg13, Simion17, Paterson20, Johnson22}, and they have been critical to discovering details about the Galactic bulge's rotating bar \citep{Binney91, Stanek94, Wegg13} along with its X-shaped morphology \citep{Weiland94, Nataf10, McWilliam10, Wegg13}.

As the nearest example of a spheroidal/bar population, and near enough to resolve individual stars, the Galactic bulge offers the possibility to explore the structure and formation history by an analysis of the spatial variation of abundances and kinematics, often called "chemodynamics".  This effort has historically required the correlation of spectroscopically derived metallicities and radial velocities, beginning with \citet{Rich90} and \citet{Minniti10}, early works that showed a trend of declining velocity dispersion with increasing metallicity. Spectroscopic surveys have grown in size from $\sim$ 10,000 stars in the Bulge Radial Velocity Assay \citep[BRAVA]{Rich07,Kunder12} to tens of thousands of stars with abundances and kinematics in projects such as the Abundances and Radial velocity Galactic Origins Survey \citep[ARGOS]{Freeman13}, the GIRAFFE Inner Bulge Survey \citep[GIBS]{Zoccali14}, and the Apache Point Observatory Galactic Evolution Experiment \citep[APOGEE]{Majewski17}.

However, even sample sizes of $10^4$ or greater can be insufficient when binned by spatial location, kinematics, and abundance. Two revolutionary developments- wide field imagers like the Dark Energy Camera on the Blanco 4m telescope \citep{Flaugher15} and the {\Gaia} astrometric survey \citep{GaiaDR3} have combined to expand sample sizes into the millions. A significant recent breakthrough has been the development of a robust photometric metallicity scale, $(u-i)_0$ versus $\rm \ [Fe/H]$ for bulge red clump (RC) giants \citep{Johnson20}, by correlating the photometry with [Fe/H] from GIBS survey spectroscopy \citep{Zoccali17}.  Combining photometric metallicities and distances for the RC with {\Gaia} kinematics potentially unlocks the exploration of chemodynamics for millions of stars, albeit with only [Fe/H] to 0.2 dex precision, and no detailed abundances (e.g., alpha elements).  Here we exploit the large numbers possible by combining the two datasets, reaching numerical sizes that are of the same order of those of the two planned major bulge surveys by 4MOST and MOONS \citep{Bensby19,Gonzalez20}.

Many investigations have pointed toward a significant break in kinematics at [Fe/H] $\sim -0.5$, beginning with \citet{Zhao94} and continuing with \citet{Soto07}, \citet{Zoccali08}, and \citet{Ness13_kinematic}.  The break at $-0.5$ dex is clearly seen when the vertex deviation is derived (this requires the construction of the velocity ellipsoid from radial velocities and longitudinal proper motions, and is interpreted as due to the metal-rich population being in the bar). \citet{Babusiaux10} explored Baade's window and two other fields on the bulge minor axis, combining proper motions from OGLE-II \citep{Sumi+04} with radial velocities and metallicities from FLAMES/GIRAFFE at the VLT. They identified two distinct populations: metal-rich stars have bar-like kinematics, while the metal-poor ones are consistent with an old spheroid (or thick disk). This supported the co-existence of both a classical and a pseudo-bulge in the central region of our Galaxy \citep[see][]{Kormendy04}. Subsequent studies have confirmed the different behavior of metal-poor and metal-rich stars in the inner Milky Way \citep[see e.g.][]{Hill11, Dekany13, Ness13_mdf, Rojas14, Babusiaux16, Kunder16, Athanassoula17, Rojas17, Zoccali17, Clarkson18, Zoccali+18, Arentsen20, Du+20, Rojas20, Simion21, Wylie21, Rix+22}. Stellar populations with different chemical compositions will also show different kinematics due to the presence of the bar, a process called kinematic fractionation \citep{Debattista17}.
Our investigation lacks radial velocities, but we are able to explore proper motions and proper motion dispersion versus [Fe/H] over an unprecedented scale, allowing us to study the dependence of kinematics on the metallicity across the spatial extent of the Southern Galactic bulge.  Future investigations will benefit from large numbers of stars with spectroscopic abundances and radial velocities, but our work aims to offer some insight into what we can expect from these surveys.

There is good reason to suspect that strong chemodynamic correlations exist in our bulge dataset.  \citet{Johnson22} shows a striking trend in which RC giants of increasing metallicity are more concentrated to the Galactic plane; there is no such trend in the radial direction.  This striking vertical abundance gradient that steepens at [Fe/H]$>-0.5$ is consistent with a complex bar with properties that depend on metallicity.  

The age of the bulge, especially of the metal-rich population, has also been called into question.  Early HST-based investigations used proper motion cleaning to find a compact old main sequence turnoff \citep{Ortolani93,Kuijken02, clarkson08,clarkson11}; these studies have been segregated by metallicity \citep{Renzini18}.   The apparently old age was confirmed over the wider bulge field population \citep[e.g.,][]{Zoccali03,Valenti13, Surot19}.  While \citet{Haywood16} have proposed that a bulge population with a complex age distribution might fit a more vertical turnoff, the bulge turnoff morphology has been investigated in multiple fields \citep[e.g.,][]{Valenti13} and does not exhibit the vertical structure required in this study.  

A more significant case for young populations is raised by \citet{Bensby17}, who derive spectroscopic ages for microlensed dwarfs in the bulge, finding a large fraction of the suprasolar metallicity stars in the bulge to be $<$5 Gyr in age.  A new analysis using MIST isochrones by \citet{Joyce+23} 
suggests the presence of a significantly smaller, but not absent, population of metal-rich stars with ages $\le8$Gyr,
but largely supports a bulge age more tightly clustered around 10-11 Gyr. Though this latter analysis, based on the same sample of microlensed dwarfs presented in \citet{Bensby17}, is in better agreement with HST-based ages, the tension between the microlening-derived ages and the HST-/main sequence turnoff-derived ages 
is not fully resolved.  

If a substantial fraction of the bar is $<$ 10 Gyr (younger than the thick disk) then one might expect the younger ages to be associated with the formation in the disk and to present a chemodynamic correlation that arises from age as well as metallicity.  At present, we lack age constraints for individual RC stars, though a rough mapping between [Fe/H] and age can be constructed from the works of \citet{Bensby17} and \citet{Joyce+23}, but our study provides metallicities that can be correlated with kinematics.

We can now build on this legacy to investigate the detailed relationship between stellar metallicity and kinematics by constructing a comprehensive sample of RC stars residing within the Milky Way bulge that combines the photometric metallicities and distances for $\sim$ 2.6 million RC stars from \citet{Johnson22} with the high-quality proper motions from the third data release of {\Gaia} \citep[DR3,][]{GaiaDR3, Lindegren+21a}.  The paper is organized as follows. In Section \ref{sec:data} we introduce the sample of RC stars with BDBS photometry, distances, and metallicities and describe the process of assigning {\Gaia} DR3 proper motions and projected velocities, that will be used in this work. In Section \ref{sec:results} we investigate the kinematics of RC stars as a function of metallicity, presenting our results, and highlighting the difference between the motion of the metal-poor and metal-rich components. Finally, in Section \ref{sec:discuss_concl} we discuss our findings, and in we summarize our results.

\section{The Red Clump Star Sample}
\label{sec:data}

\begin{figure}[]
\centering
\includegraphics[width=\columnwidth]{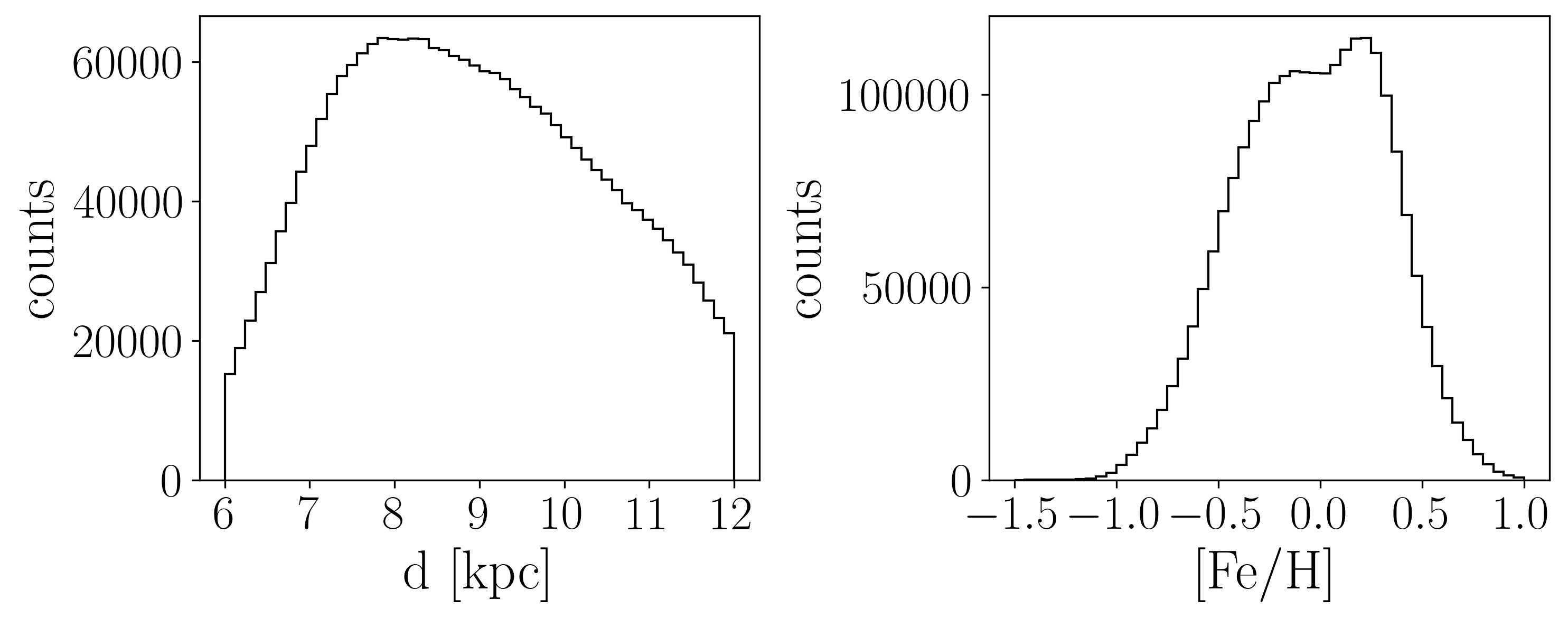}
\caption{Distance (left panel) and metallicity (right panel) distributions of the sample of $2.6$ million RC stars, as derived from BDBS photometry in \citet{Johnson20, Johnson22}.}
\label{fig:hist_met}
\end{figure}

\begin{figure*}[]
\centering
\includegraphics[width=\textwidth]{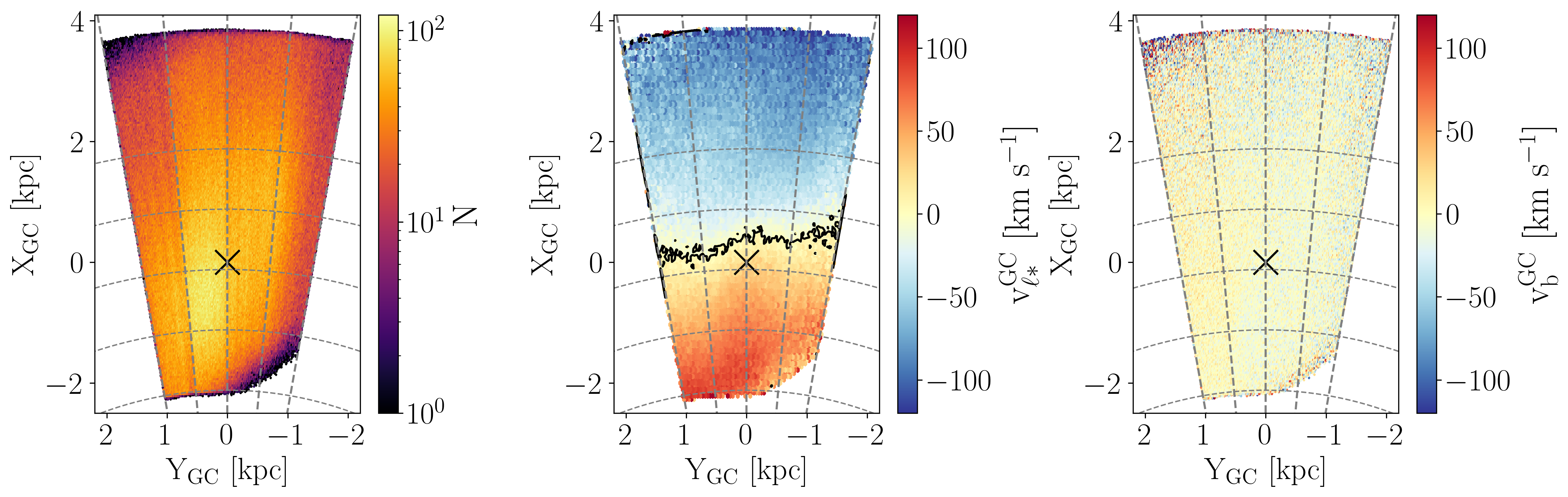}
\caption{Top-down views of the RC stars in Galactocentric Cartesian coordinates. The black cross marks the position of the Galactic Centre, at $(X_\mathrm{GC}, Y_\mathrm{GC}) = (0, 0)$. Grey dashed lines correspond to constant values of Galactic longitude ($\ell = -10^\circ, -5^\circ, 0^\circ, 5^\circ, 10^\circ$), and grey dashed arcs correspond to constant heliocentric distances of $6$, $7$, $8$, $9$, and $10$ kpc. Left: Logarithmic density of stars in each bin of position. Middle: Galactic longitudinal velocity. The black contour corresponds to the line of nodes $v_{\ell*}^\mathrm{GC}=0$. Right: Galactic latitudinal velocity.}
\label{fig:RC_xy_vlvb}
\end{figure*}

\begin{figure}[]
\centering
\includegraphics[width=0.6\columnwidth]{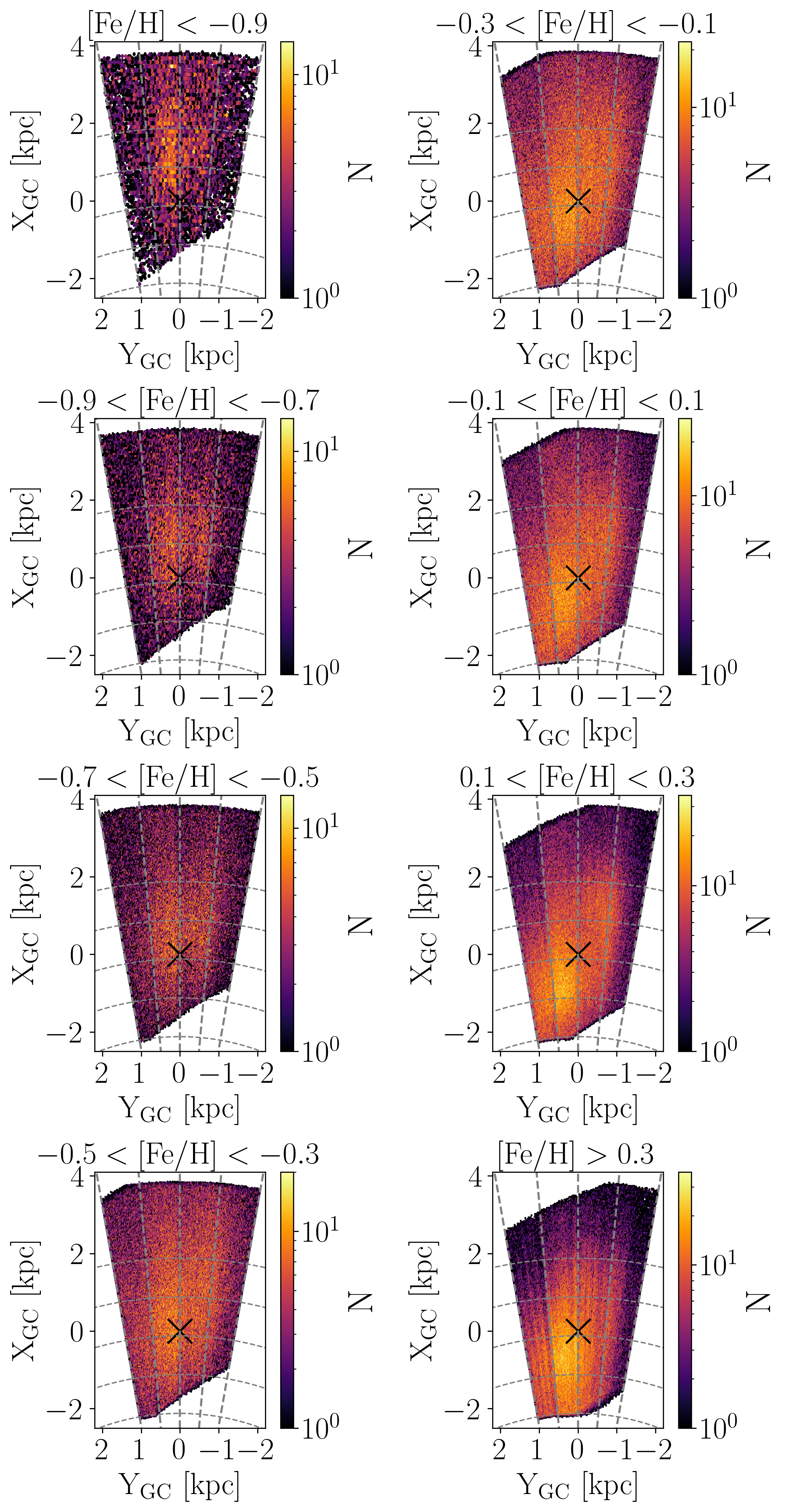}
\caption{Same as the Fig.\ref{fig:RC_xy_vlvb} (left panel), but for the metallicity bins used in Fig. 19 of \citet{Johnson22}.}
\label{fig:XY_met}
\end{figure}

In this work, we start with the sample of $2.6$ million RC stars extracted from BDBS data in \citet{Johnson20, Johnson22}, spanning the region in Galactic coordinates $(\ell, b)$ $-10^\circ \leq \ell \leq 10^\circ$, $-10^\circ \leq b \lesssim -3^\circ$ of the southern Galactic bulge\footnote{An electronic table containing the positions, $ugi$-band photometry, extinction, distance, and [Fe/H] values for all stars used here is provided in Table 1 of \citet{Johnson22}.}. Thanks to the sensitivity of the near-ultraviolet $u$ band to the metallicity of the RC stars, \citet{Johnson20, Johnson22} derived photometric metallicities for the whole sample, with a typical dispersion of $0.2$ dex, employing the dust map constructed in \citet{Simion17} using infrared VVV data. The distance distribution of the RC sample, derived using BDBS photometry \citep{Johnson20, Johnson22}, is shown in the left panel of Fig. \ref{fig:hist_met}. RC distances span a range of $[6, 12]$ kpc, with a median value of $8.83$ kpc. The metallicity of the sample, shown in the right panel of Fig. \ref{fig:hist_met}, covers a range going from $-1.5$ to $1$ dex, with a median value of {\Feh} $\sim -0.04$ dex. The two peaks of the metallicity distribution are at {\Feh} $\sim -0.15$ dex and {\Feh} $\sim 0.2$ dex. More information on the metallicity distribution function and on the spatial distribution of the RC stars is provided in \citet{Johnson22}, including a description of the red clump color and magnitude selection functions (see their Section 2).

We cross-match this dataset to {\Gaia} DR3 data, back-propagating {\Gaia} coordinates to the mean BDBS epoch through {\Gaia} proper motions, using a tolerance radius of 1 arcsec \citep[see][for further details on the cross-matching procedure]{Marchetti+22}. This results in a sample of $2593172$ RC stars with BDBS photometry (and therefore photometric distances and metallicities), and {\Gaia} DR3 astrometry (parallaxes and proper motions). We further select sources with a value of the {\Gaia} DR3 Renormalised Unit Weight Error (RUWE) $<1.4$, to avoid contamination from spurious astrometric solutions \citep{Lindegren+21a, Belokurov20, Penoyre20}. After this selection cut, we are left with a sample of $2315197$ RC stars, which will be the main focus of this paper. Thanks to the homogeneous {\Gaia} DR3 proper motions and BDBS metallicities over the large BDBS instrumental footprint, this is an ideal dataset to study the chemo-kinematics of RC stars across the whole southern Galactic bulge. 

While random and systematic uncertainties in {\Gaia} parallaxes are too large to provide reliable distances for individual stars in the Galactic bulge \citep[e.g.][]{Lindegren+21}, we rely on photometric distances and {\Gaia} DR3 proper motions to investigate the kinematics of the RC sample. 
We convert Galactic proper motions $(\mu_{\ell*} \equiv \mu_{\ell}\cos b, \mu_b)$ to Galactocentric velocities along longitude $v_{\ell*}^{\mathrm{GC}}$ and latitude $v_b^{\mathrm{GC}}$ by subtracting the contribution given by the motion of the Sun:

\begin{equation}
    \label{eq:vl_GC}
    v_{\ell *}^{\mathrm{GC}} = v_{\ell *} - U_\odot\sin l\cos b + V_\odot \cos l \cos b \ ,
\end{equation}
\begin{equation}
    \label{eq:vb_GC}
    v_b^{\mathrm{GC}} = v_{b} + W_\odot \cos b ,
\end{equation}
where $v_{i} \ \mathrm{[km/s]} = 4.74 \cdot \mu_{i} \mathrm{[mas/yr]} \cdot d \mathrm{[kpc]}$ for $i = \ell*, b$, and $d$ is the heliocentric distance of the star, derived with BDBS photometry \citep{Johnson20}. We assume a three-dimensional Cartesian velocity of the Sun $[U_\odot, V_\odot, W_\odot] = [12.9, 245.6, 7.78]$ km s$^{-1}$ \citep{Reid04, Drimmel+18, Gravity18}.

In the left panel of Fig. \ref{fig:RC_xy_vlvb}, we show the distribution of the sample of RC stars in Cartesian Galactocentric coordinates $(X_\mathrm{GC}, Y_\mathrm{GC})$, where the black cross marks the position of the Galactic Centre, assuming a distance of $8.122$ kpc \citep{Gravity18}. The $X_\mathrm{GC}$ coordinate is aligned to the Sun-Galactic Centre direction, and is positive in the direction of the Galactic Centre, while the $Y_\mathrm{GC}$ coordinate is positive along the direction of the Sun circular rotation in the disk. As already shown by previous works \citep[e.g.][]{Wegg13, Johnson22}, the over-density at $Y_\mathrm{GC} > 0$, $X_\mathrm{GC} < 0$ is due to the orientation of the near side of the bar, forming an angle of $\sim 27^\circ$ with the line of sight at $\ell=0$ \citep{Wegg13}.
If we split our sample in metallicity, using the same bins in [Fe/H] adopted by \citet{Johnson22}, Fig.\ref{fig:XY_met} shows that the distribution of the stars is heavily dependent on their chemical composition. Metal-poor stars ([Fe/H]$\lesssim -0.5$ dex) are more centrally distributed around the Galactic Centre, while the morphology of more metal-rich stars ([Fe/H]$\gtrsim -0.5$ dex) shows the typical asymmetry with Galactic longitude caused by the orientation of the bar.

\section{Chemo-kinematics of the Red Clump Stars}
\label{sec:results}

In this Section, we inspect the projected kinematics of the RC stars, as a function of position and metallicity, in the southern Galactic bulge.  However, we caution that the spatial distributions presented here are heavily affected by our $\sim$10-20$\%$ distance uncertainties.  As Figure 8 of \citet{Hey23} shows, distance uncertainties of this magnitude distort two and three dimensional projections, particularly on the near side of the Galactic Center, and can smear out otherwise well-defined features.

In the central and right panel of Fig. \ref{fig:RC_xy_vlvb},  we plot the distribution of RC stars color-coded by Galactocentric velocity along Galactic longitude and latitude, respectively (equations \ref{eq:vl_GC} and \ref{eq:vb_GC}). The clear asymmetry in the velocity field along longitude with respect to $Y_\mathrm{GC}$ is an indication of the presence of the bar, which breaks the axial symmetry of the potential in the inner Galaxy \citep[see also][]{Sanders19}. The orientation of the $v_{\ell*}^{\mathrm{GC}} =0$ contour line is $\sim 80^\circ$ at $Y_\mathrm{GC}=0$, consistently with results from \citet{Sanders19}. As expected, the mean velocity along latitude is $v_b^{\mathrm{GC}}\sim 0$, indicating no clear streaming motion above/below the Galactic plane \citep[e.g.][]{Reid04, Du+20}.

\subsection{Proper motion rotation curves as a function of metallicity}
\label{sec:rot_curves}

\begin{figure}[]
\centering
\includegraphics[width=0.7\columnwidth]{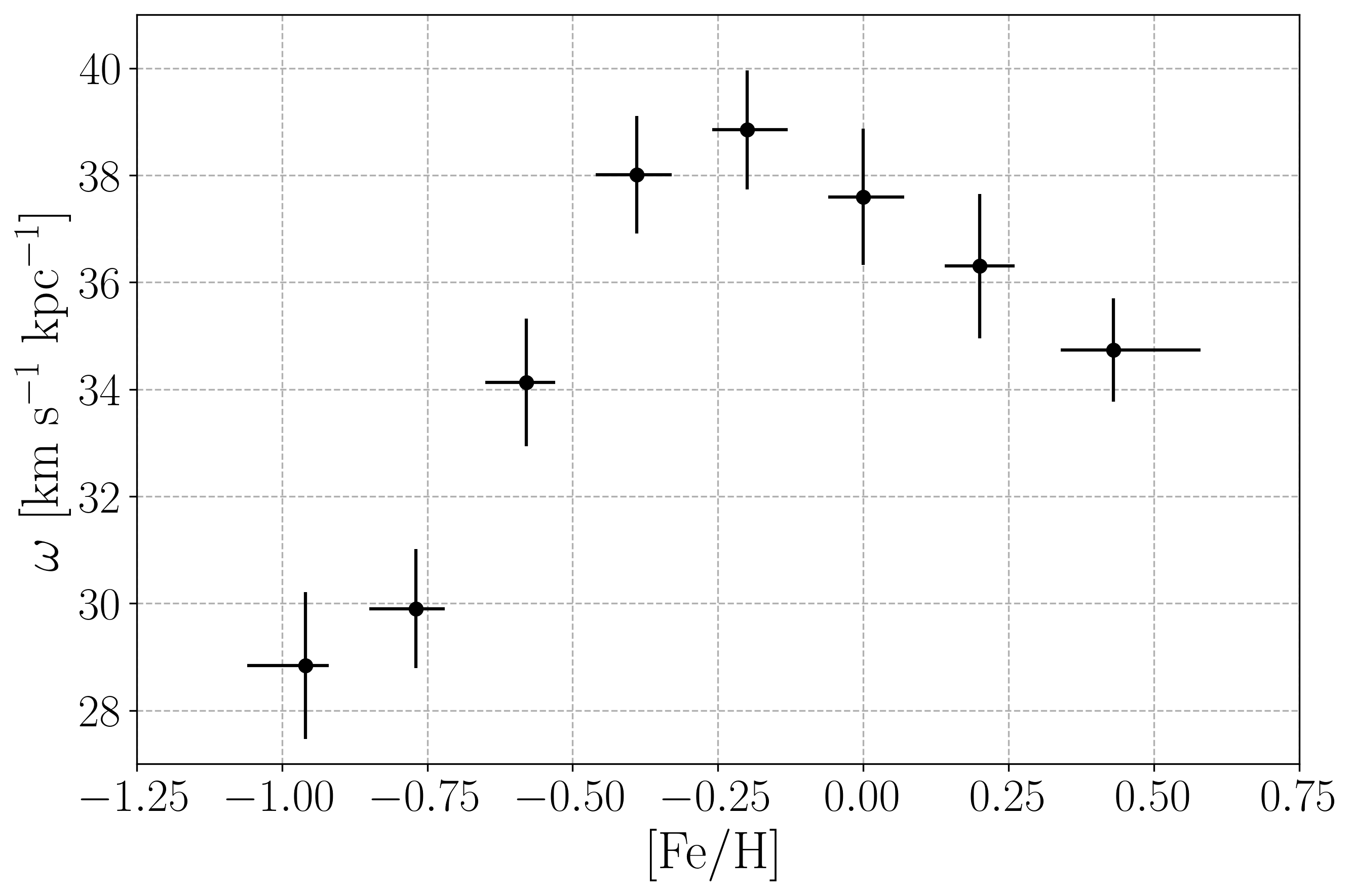}
\includegraphics[width=0.7\columnwidth]{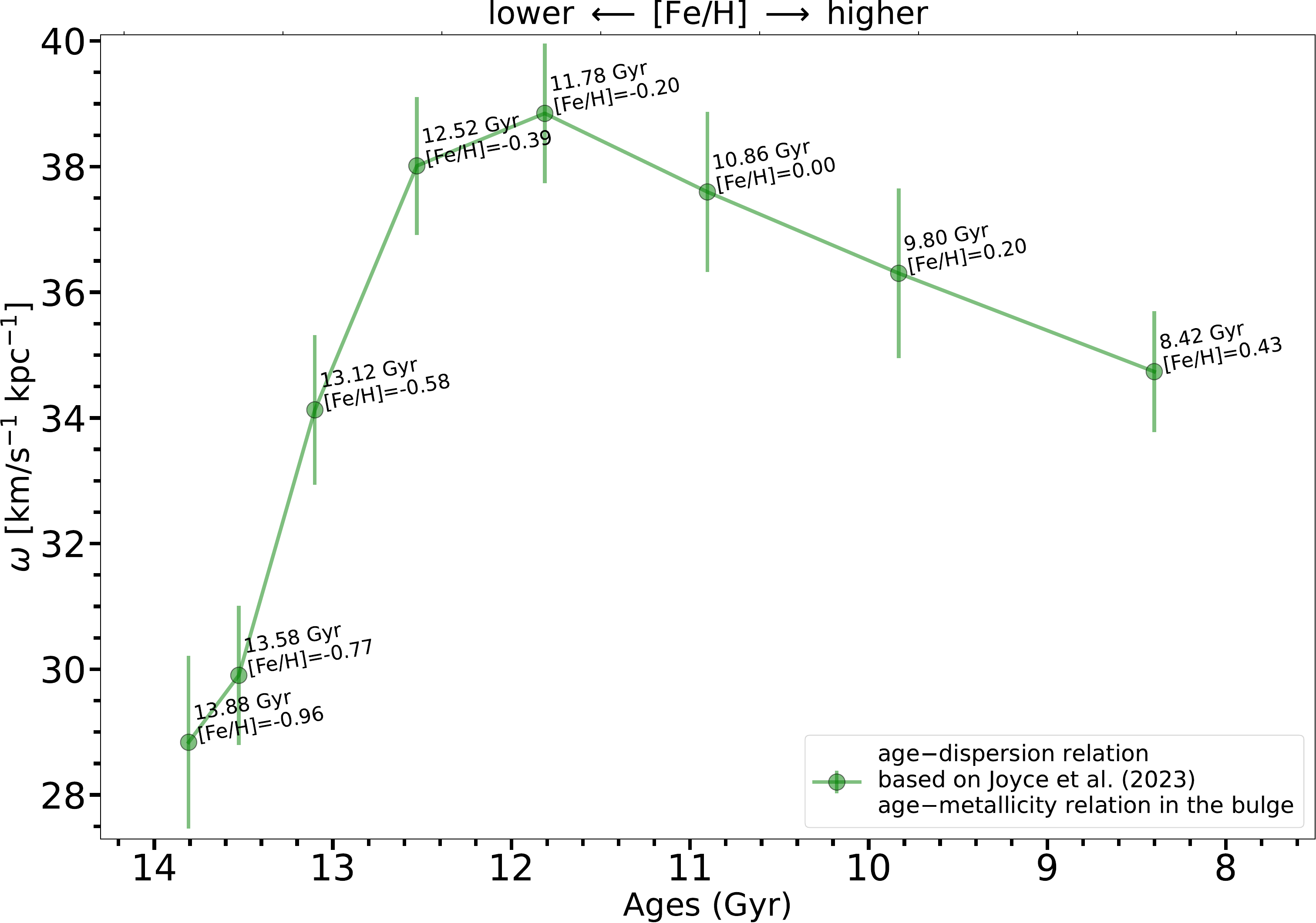}
\caption{Upper: Angular velocity $\omega$ as a function of metallicity for the whole sample of RC stars. Uncertainties on the angular velocity are derived by the fitting procedure, while the ones on the x-axis correspond to the 16$^\mathrm{th}$ and 84$^\mathrm{th}$ percentiles of the metallicity distribution in each bin. 
Lower: Ages inferred from the age--metallicity relation described in \citet{Joyce+23} are assigned to the metallicity bins shown above, yielding angular velocity as a function of age. The x-axis is inverted (e.g. ages shown in reverse) for better visual alignment with the panel above.}
\label{fig:angular_v_feh}
\end{figure}

\begin{figure}[]
\centering
\includegraphics[width=\columnwidth]{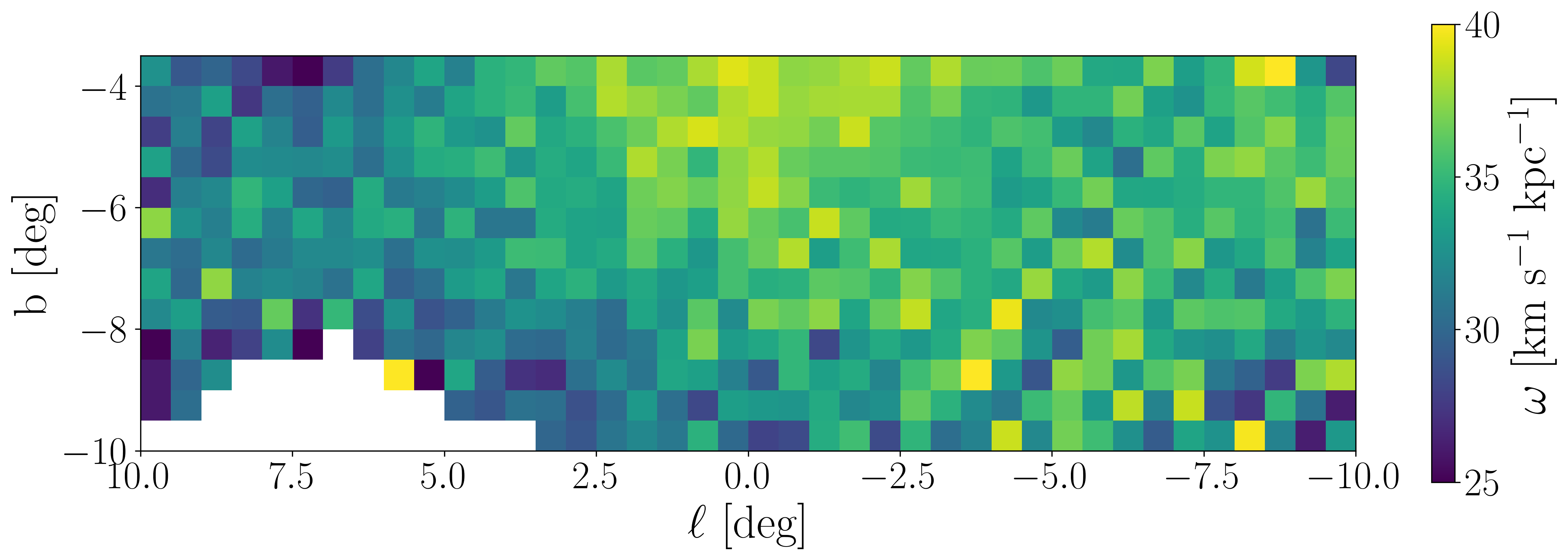}
\caption{Angular velocity $\omega$ as a function of Galactic coordinates in the southern Galactic bulge region, computed using all the RC stars in our sample. The bins have sizes of $0.5^\circ\times0.5^\circ$.}
\label{fig:angular_v_2d}
\end{figure}

\begin{figure*}[]
\centering
\includegraphics[width=\textwidth]{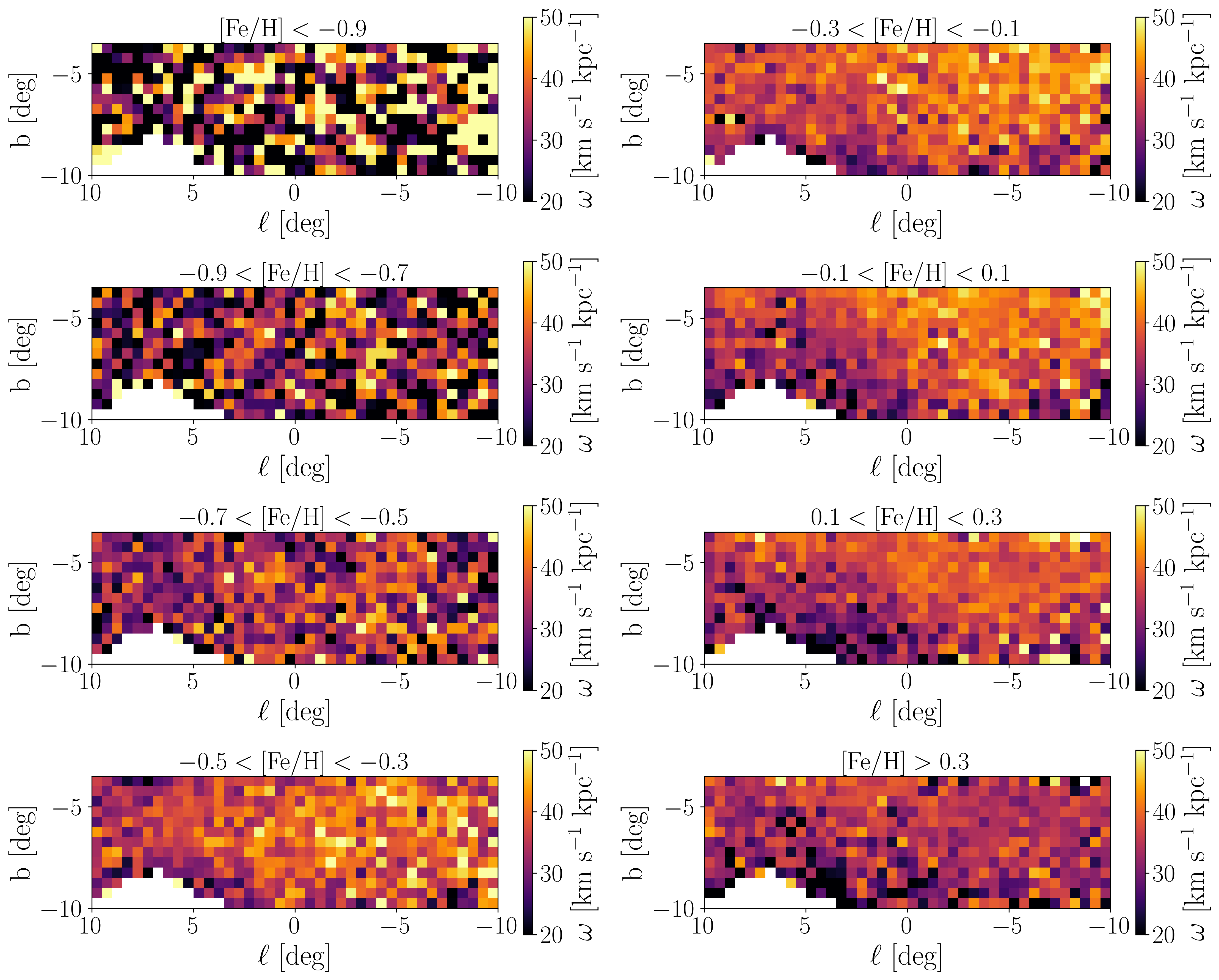}
\caption{Angular velocity $\omega$ as a function of Galactic coordinates, for different metallicities of the RC sample. The metallicity bins are the same as in Fig.\ref{fig:XY_met}.}
\label{fig:angular_v_2d_feh}
\end{figure*}

{\Gaia} DR3 proper motions can be further used to construct proper motion rotation curves, as a function of heliocentric distance. Longitudinal Galactic proper motions provide similar results to line-of-sight velocities in the direction of the Galactic bulge \citep{Du+20}. By plotting the median of the Galactic longitudinal velocity (as given by Eq.\eqref{eq:vl_GC}) as a function of distance $d$, we can compute the angular velocity $\omega$ as the slope of the linear relation \citep[e.g.][]{Du+20}, for different metallicity bins. By binning the stars in metallicity using the same intervals adopted by \citet{Johnson22}, in the upper panel of Fig. \ref{fig:angular_v_feh} we show the resulting $\omega$ as a function of {\Feh} (we report the median metallicity value in each bin) for our sample of RC stars. The uncertainty on $\omega$ is computed from the linear fit using the method of least-squares, taking into account the uncertainties on the velocity (computed by dividing the velocity dispersion by the square root of the number of stars in each metallicity bin). 
Metal-poor RC stars rotate slower: $\omega \sim 29$ km s$^{-1}$ kpc$^{-1}$, and then the angular velocity increases significantly at {\Feh} $\sim -0.5$ dex, with a maximum value $\omega = 39$ km s$^{-1}$ kpc$^{-1}$ for {\Feh} $\sim -0.2$ dex. The angular velocity starts decreasing again for {\Feh} $\gtrsim -0.2$ dex, reaching a value of $\sim 35$ km s$^{-1}$ kpc$^{-1}$ in the most metal-rich bin at {\Feh} $\gtrsim 0.4$ dex. The lower rotation of metal-poor stars is consistent with several studies of the Galactic bulge \citep[e.g.][]{Ness13_kinematic, Ness16, Zoccali17, Clarkson18}. 

We can now use the age--metallicity relation presented in \citet{Joyce+23} to roughly estimate the dependence of the angular velocity versus the age of the stars. This is shown in the lower panel of Fig.\ref{fig:angular_v_feh}
We do not attempt to compute proper age uncertainties for this analysis, so no horizontal error bars are given (conservatively, one expects uncertainties of roughly $1.5- 2$ Gyr; \citealt{Joyce+23}). As expected by the monotonic behavior of the adopted relation, the shape of the curve reflects the chemical one. The peak of the angular velocity corresponds to a population with an age of $\sim 11.8$ Gyr. The oldest stars in our sample have ages comparable to the age of the Universe, while the metal-rich stars are older than 8 Gyr.

Thanks to the large spatial coverage of our sample in the southern Galactic bulge, and to the availability of homogeneous all-sky {\Gaia} DR3 proper motions, in Fig. \ref{fig:angular_v_2d} we plot the sky distribution in Galactic coordinates of the angular velocity $\omega$, for RC stars of all metallicities. We see that $\omega$ peaks at $\sim 40$ km s$^{-1}$ kpc$^{-1}$ at low latitudes along the bulge minor axis, and it decreases to a minimum of $\sim 25$ km s$^{-1}$ kpc$^{-1}$ for off-axis fields at higher latitudes. Typical uncertainties from the fitting procedures are $\sim 3$ km s$^{-1}$ kpc$^{-1}$, and they do not show any significative dependence on the sky location. 

A comparison between Fig.~\ref{fig:angular_v_2d} here and Fig.~16c in \citet{Johnson22}, which shows spatial distribution differences in [Fe/H] between the metal-rich peak and half-power position of the metal-rich tail, indicates that the sky patterns are highly correlated.  Both distributions peak along the minor axis close to the plane, and show noticeable enhancements in both values within a region encompassing approximately $|\ell|$ $<$ 3$^\circ$ and $|b|$ $<$ 6$^\circ$.  This observation provides one of the few clear connections between chemistry and kinematics for bulge formation and evolution as the regions showing the highest angular velocities are also those for which the broadest high metallicity tails formed.

Fig. \ref{fig:angular_v_2d_feh} incorporates the spatial and chemical dependence of the Galactic longitudinal proper motion rotation curves, by presenting the resulting angular velocity $\omega$ as a function of Galactic coordinates, for several metallicity bins, chosen to match those in Fig. 19 of \citet{Johnson22}. At the lowest metallicities ({\Feh} $\lesssim -0.7$ dex), there is no evidence for a clear, coherent signal over the survey footprint, with a noisy distribution with values ranging from $20$ to $50$ km s$^{-1}$ kpc$^{-1}$. On the other hand, for {\Feh} $\gtrsim -0.5$ dex, there is a clear continuous pattern over the sky, with a mean value $\omega \sim 40$ km s$^{-1}$ kpc$^{-1}$. We observe a clear asymmetry with Galactic longitude, showing lower values of $\omega$ for positive $\ell$. We suspect this to be a projection effect due to the orientation of the bar. We do not observe a strong gradient with Galactic latitude. As also evident from Fig. \ref{fig:angular_v_feh}, the angular velocities across the sky decrease to $\sim 30$ km s$^{-1}$ kpc$^{-1}$ for {\Feh} $\gtrsim 0$, confirming the slower rotation of the most metal-rich population observed in Fig. \ref{fig:angular_v_feh}.

\subsection{Impact of the Galactic bar on the chemo-kinematics of red clump stars}
\label{sec:chemo_kin}

\begin{figure*}[]
\centering
\includegraphics[width=\textwidth]{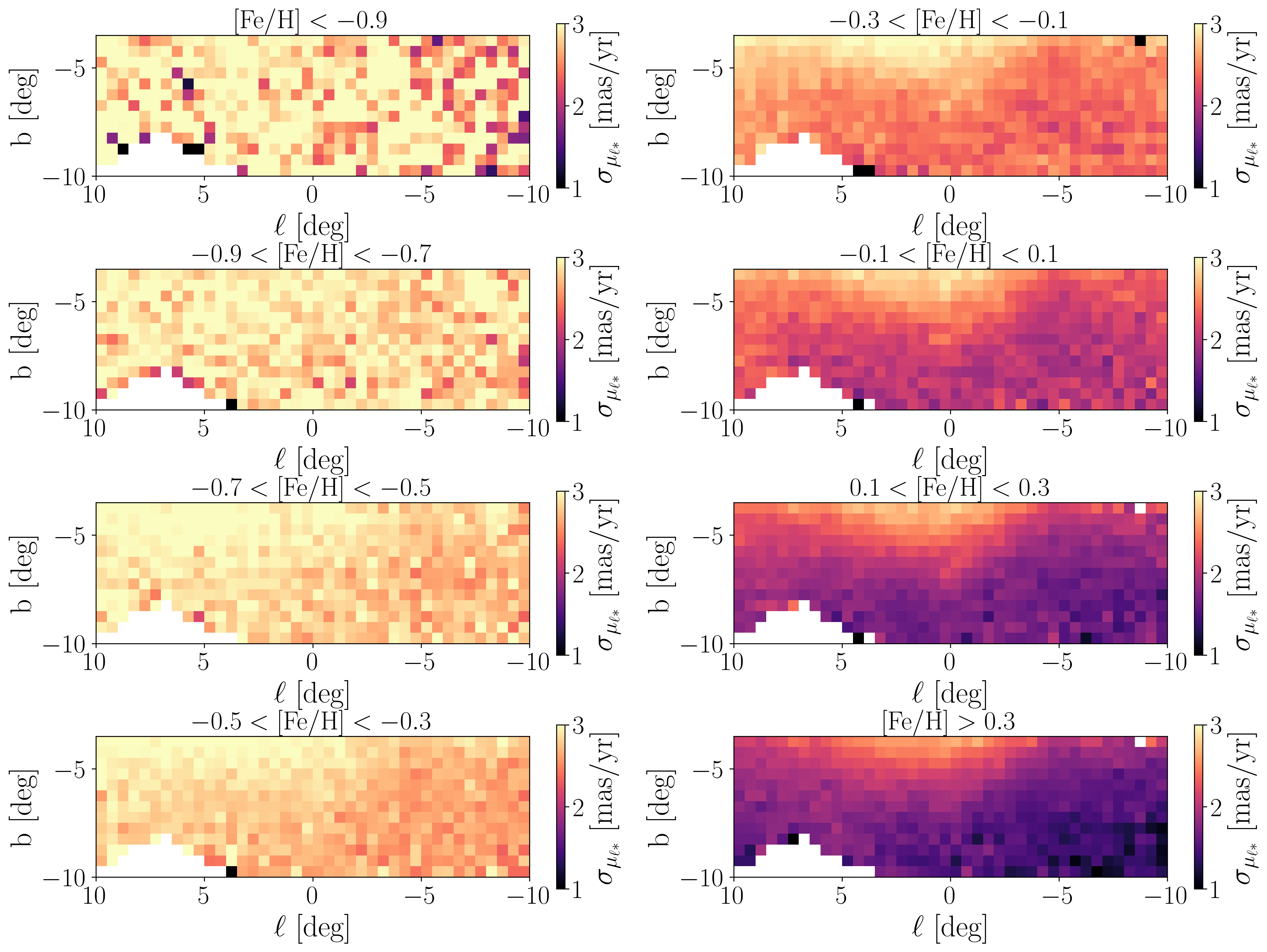}
\caption{Proper motion dispersion along Galactic longitude as a function of Galactic coordinates in the southern bulge, for the metallicity bins considered in this work and in Fig. 19 of \citet{Johnson22}.
}
\label{fig:sigmal_feh}
\end{figure*}

\begin{figure*}[]
\centering
\includegraphics[width=\textwidth]{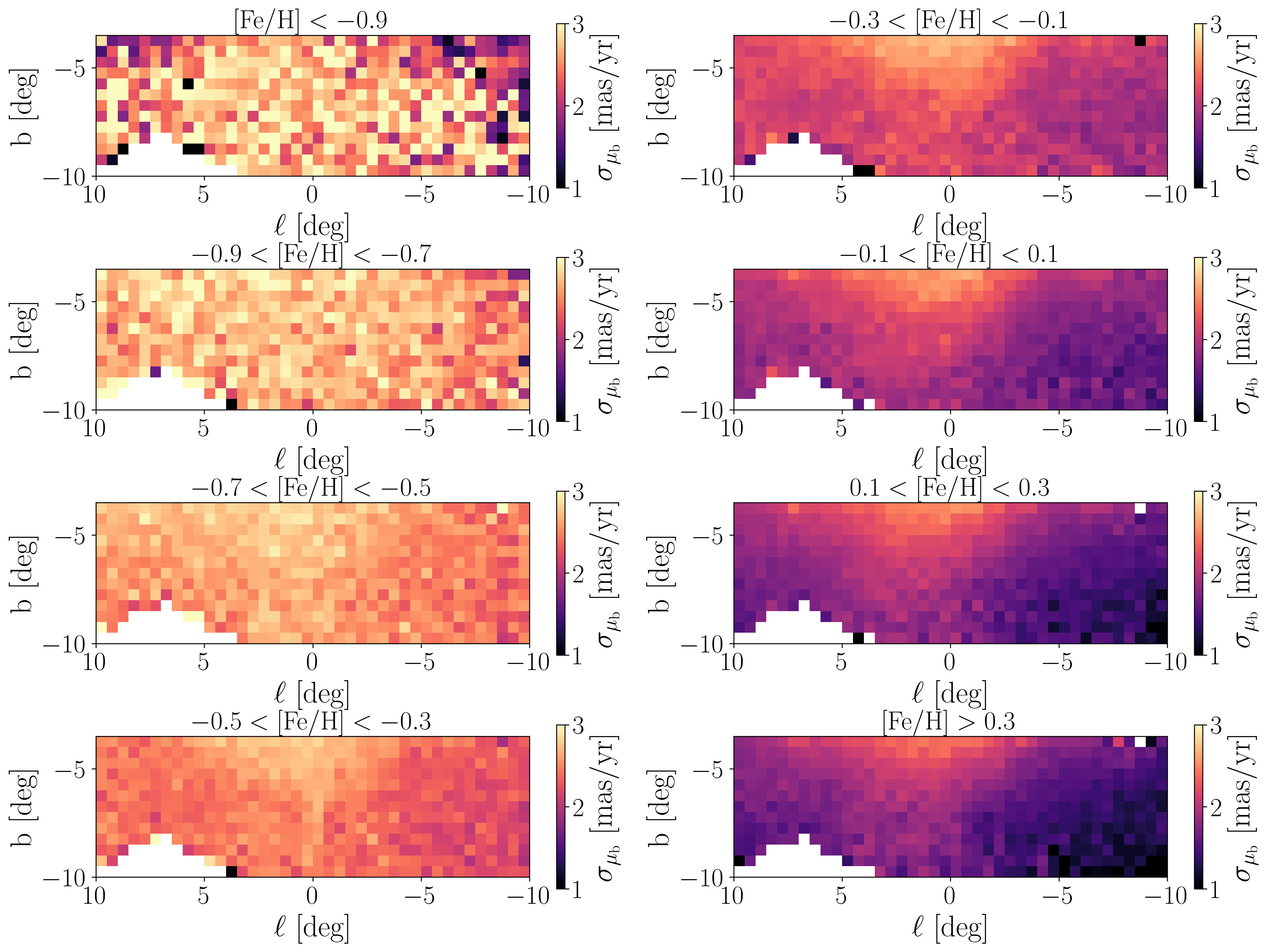}
\caption{Same as Fig.\ref{fig:sigmal_feh}, but showing the proper motion dispersion along Galactic latitude.}
\label{fig:sigmab_feh}
\end{figure*}

\begin{figure*}[]
\centering
\includegraphics[width=\textwidth]{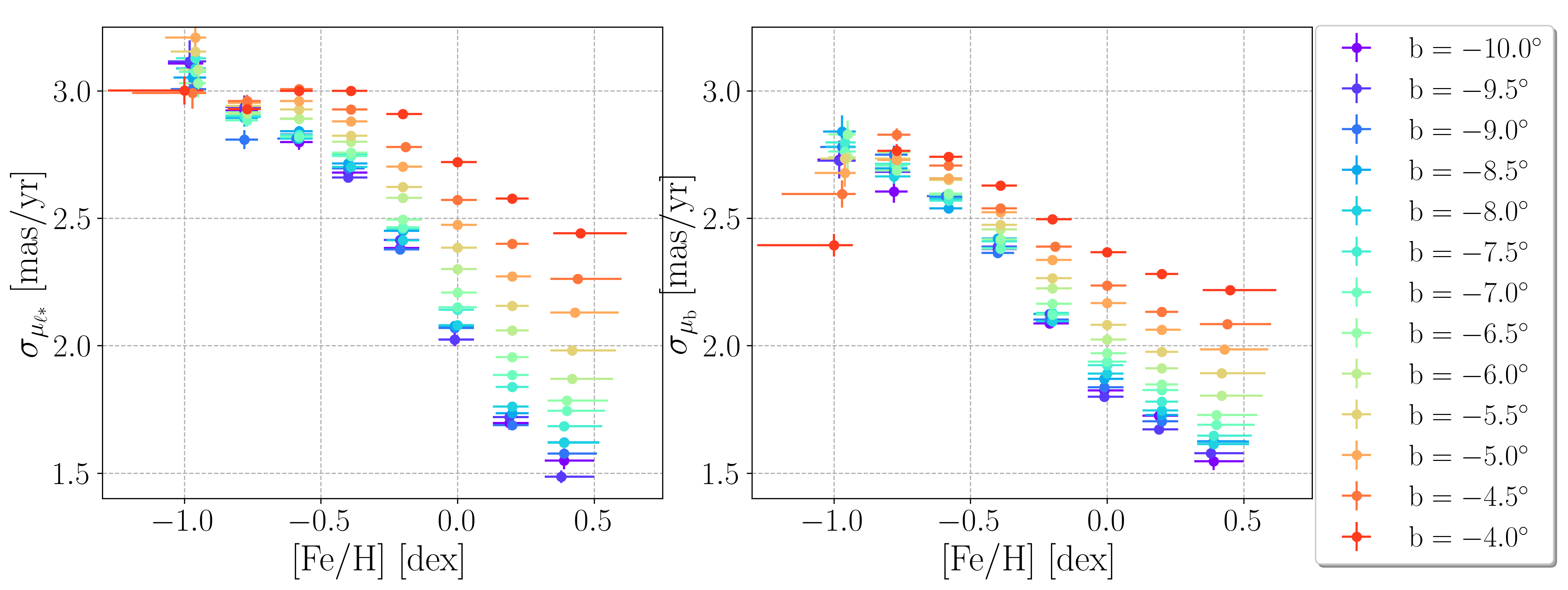}
\includegraphics[width=0.4\textwidth]{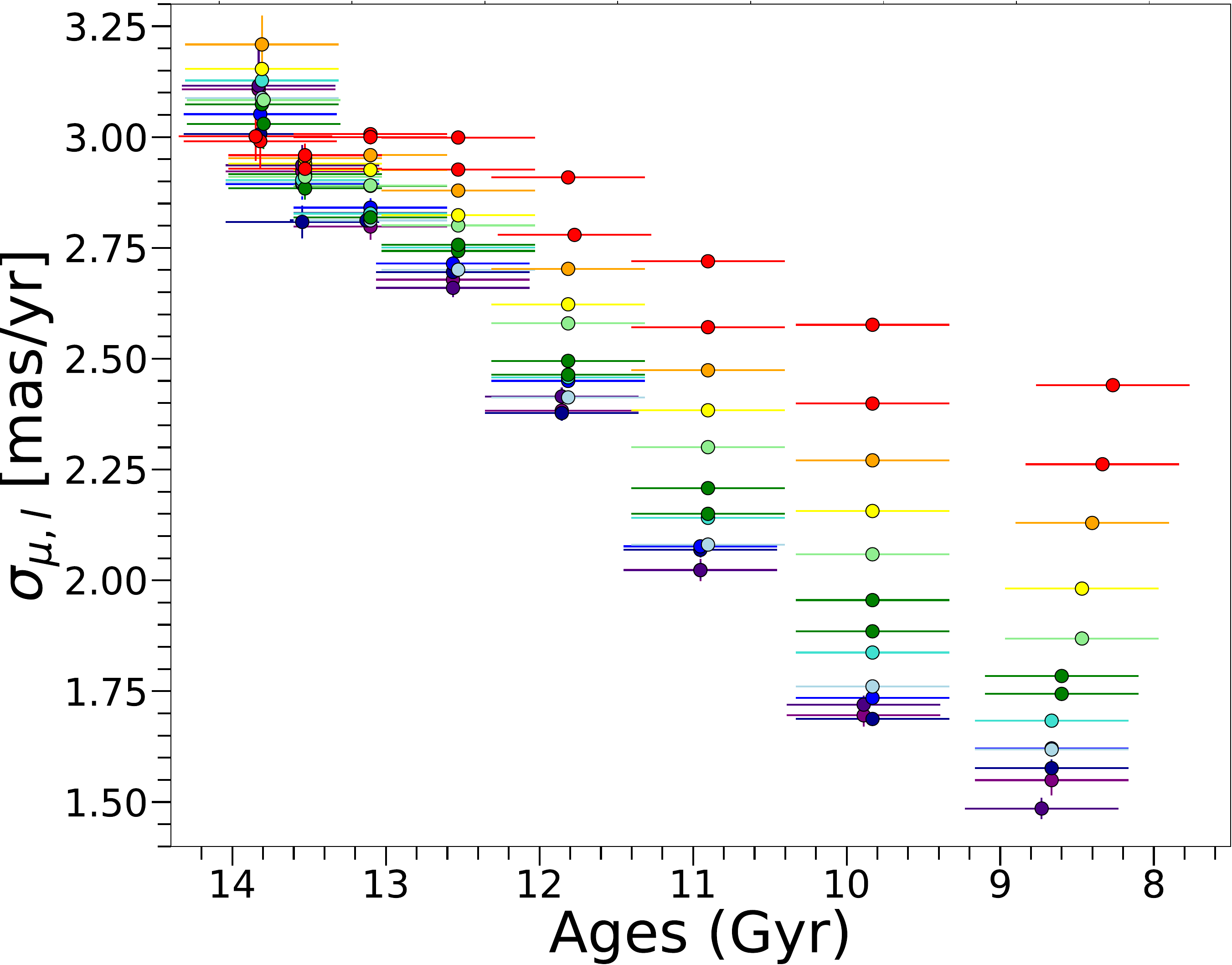} \hspace{0.3cm}
\includegraphics[width=0.4\textwidth]{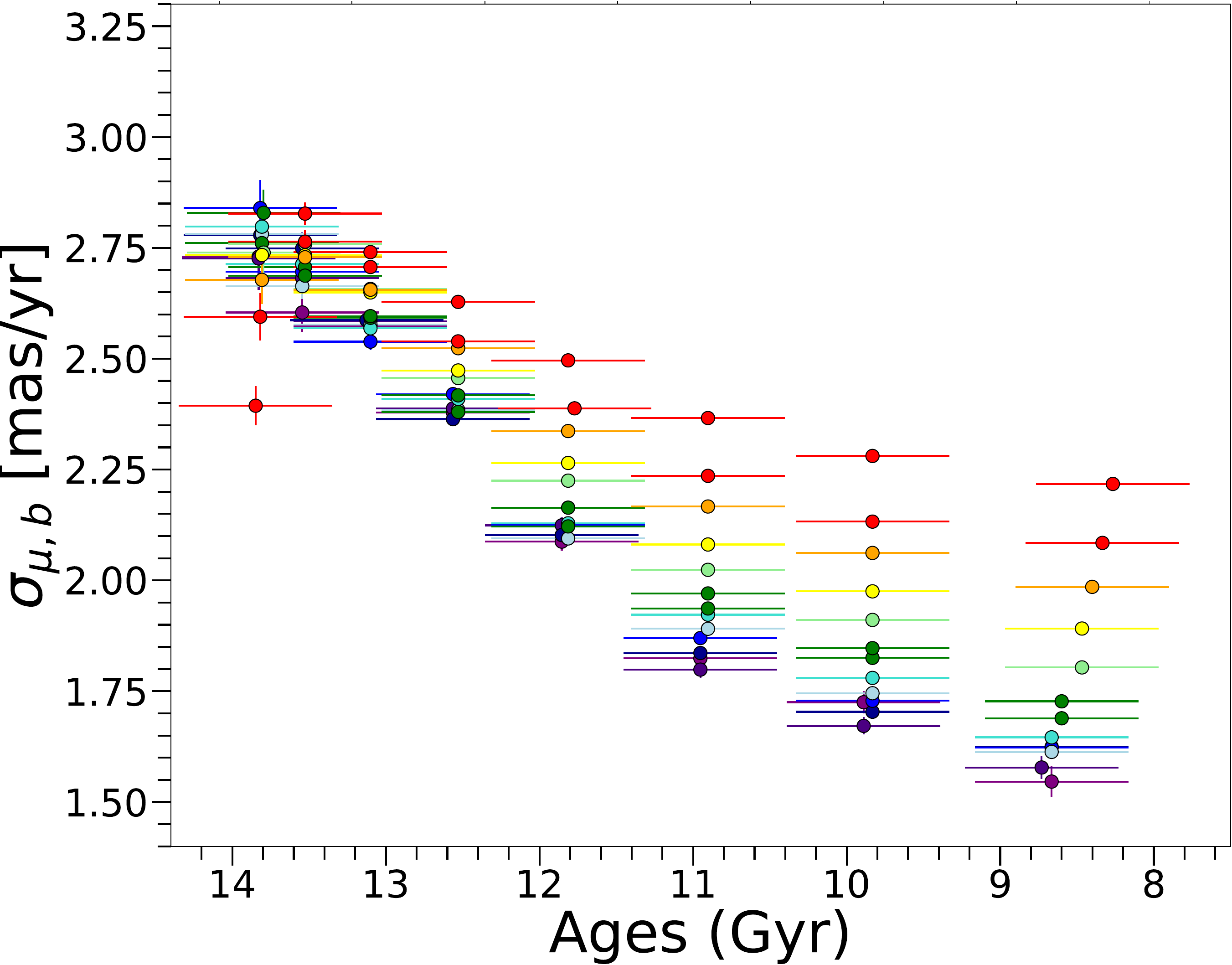} \hspace{2.5cm}
\caption{Upper: Proper motion dispersion along Galactic longitude (left panel) and Galactic latitude (right panel) as a function of metallicity, for different values of Galactic latitude.
Lower: The same, but shown with the age--metallicity substitution from \citet{Joyce+23} described in Section \ref{sec:rot_curves}.
}
\label{fig:sigmal_sigmab_feh}
\end{figure*}

\begin{figure*}[]
\centering
\includegraphics[width=\textwidth]{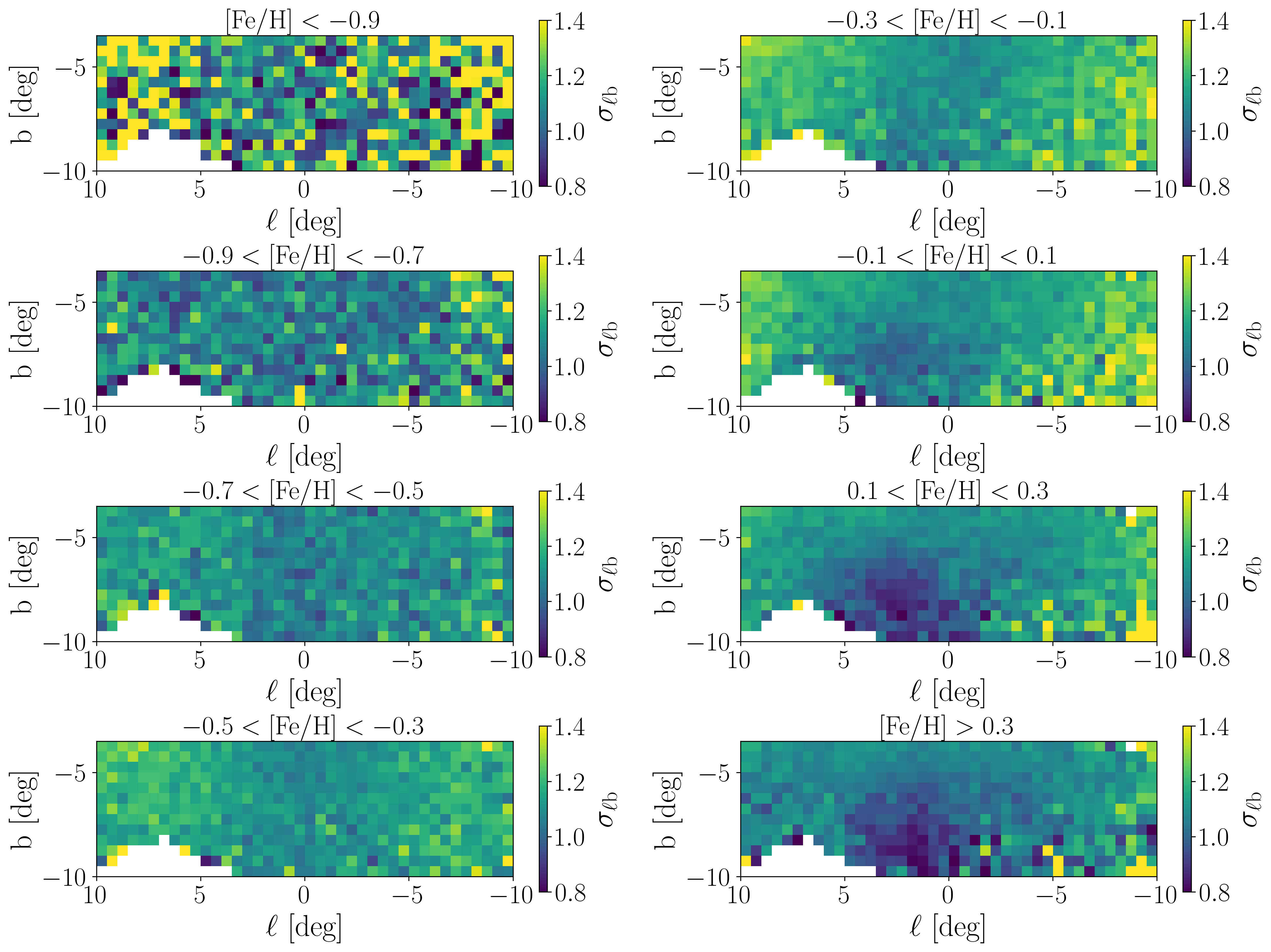}
\caption{Same as Fig.\ref{fig:sigmal_feh}, but showing the dispersion ratio, defined as the ratio between the proper motion dispersion along Galactic longitude and the one along Galactic latitude. 
}
\label{fig:dispratio_feh}
\end{figure*}

\begin{figure*}[]
\centering
\includegraphics[width=\textwidth]{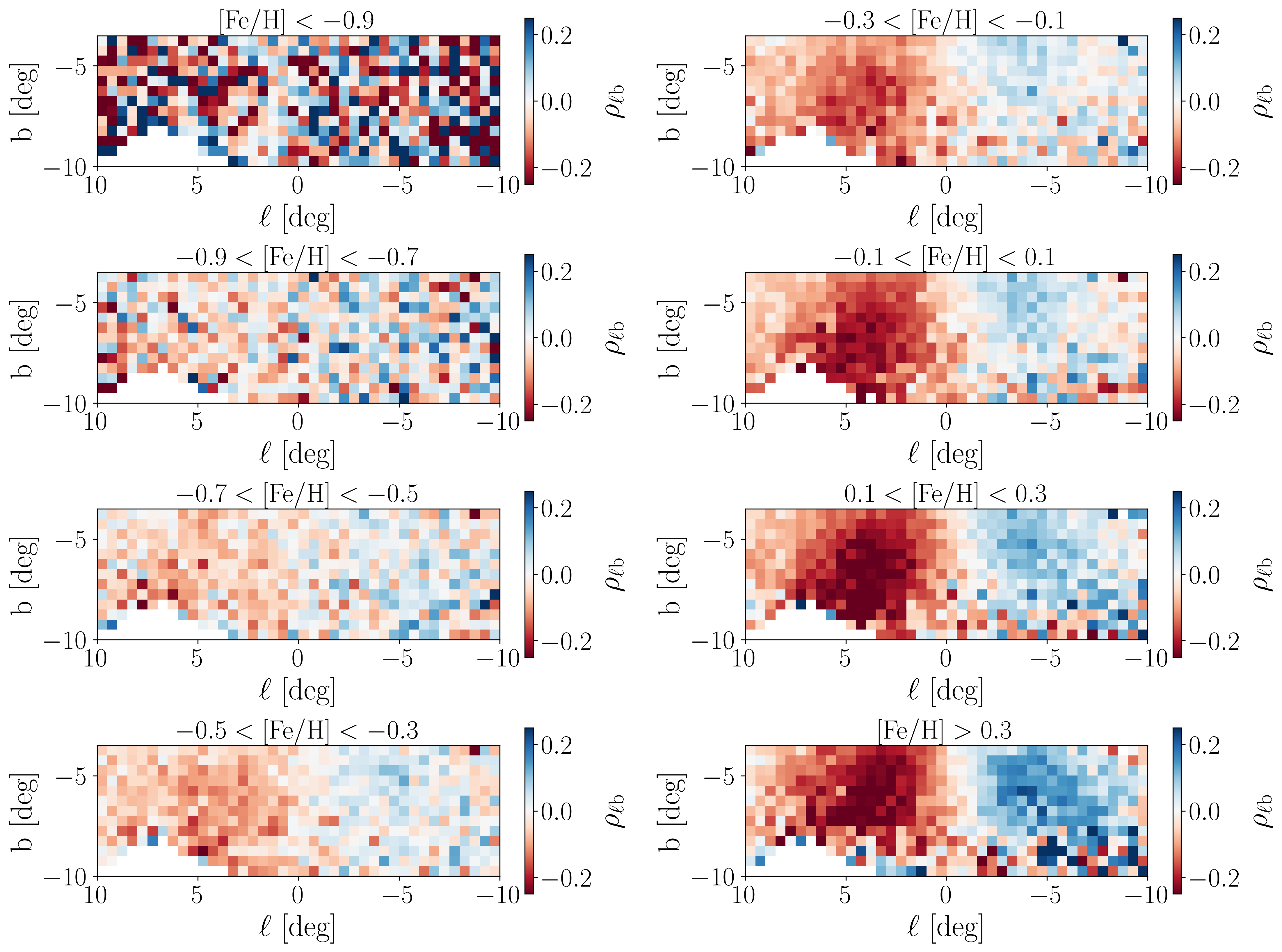}
\caption{Correlation between Galactic proper motions as a function of Galactic coordinates in the southern bulge, for the metallicity bins considered in this work, and in Fig. 19 of \citet{Johnson22}.
}
\label{fig:correlation_feh}
\end{figure*}

Recent works, combining {\Gaia} DR2 and VVV infrared data, showed the impact of the Galactic bar on the transverse kinematics of bulge stars \citep{Clarke19, Sanders19}. The authors quantified these aspects by investigating the proper motion dispersions, dispersion ratio, and correlation between Galactic proper motions as a function of Galactic coordinates across the bulge. They found higher dispersions at low latitudes, a characteristic X-shape structure in the dispersion ratio, and an approximate radial alignment in the proper motions with a clear asymmetry at positive $\ell$, due to the orientation of the near-side of the bar.

We can now reproduce these transverse kinematic maps using more precise and accurate {\Gaia} DR3 data, adding the further dimension provided by photometric metallicities. We bin the sample in metallicities using the same bins adopted by \citet{Johnson22} and in Fig. \ref{fig:angular_v_2d_feh}, and in Galactic coordinates by adopting bin sizes of $0.5^\circ \times 0.5^\circ$. In each bin, we then compute the Galactic longitudinal proper motion dispersion as:
\begin{equation}
    \label{eq:pm_disp}
    \sigma_{\mu_{\ell*}}(\ell, b, [\mathrm{Fe/H}]) = \sqrt{\langle \mu_{\ell*}^2\rangle - \langle \mu_{\ell*}\rangle^2 } \ ,
\end{equation}
and the correlation between the Galactic proper motions as:
\begin{equation}
    \label{eq:pm_corr}
    \rho_{\ell, b}(\ell, b, [\mathrm{Fe/H}]) = \frac{\langle\mu_{\ell*}\mu_b\rangle - \langle\mu_{\ell*}\rangle\langle\mu_b\rangle}{\sqrt{\Bigl(\langle \mu_{\ell*}^2\rangle - \langle \mu_{\ell*}\rangle^2 \Bigr) \Bigl(\langle \mu_b^2\rangle - \langle \mu_b\rangle^2 \Bigr)}} \ .
\end{equation}

Fig. \ref{fig:sigmal_feh} shows the resulting proper motion dispersion along Galactic longitude, $\sigma_{\mu_{\ell*}}$, as a function of sky positions, for the same metallicity bins adopted in Fig. \ref{fig:angular_v_2d_feh}. We can see how the metal-poor population is kinematically hotter, $\sigma_{\mu_{\ell*}} \sim 3$ mas yr$^{-1}$, which corresponds to $\sim 115$ km s$^{-1}$ at a distance of $8$ kpc, while the metal-rich stars show a lower velocity dispersion \citep[e.g.][]{Ness13_kinematic, Athanassoula17, Zoccali+18}. In addition, stars with {\Feh} $\gtrsim -0.5$ dex exhibit a coherent pattern over the sky, reproducing results by \citet{Clarke19, Sanders19}. This is particularly evident at {\Feh} $\sim 0$, where the RC population is kinematically colder at higher latitudes, $\sigma_{\mu_{\ell*}} \sim 2$ mas yr$^{-1}$ $\sim 76$ km s$^{-1}$, and the peak is observed on the minor axis towards the centre of the Galaxy because of the Galactic potential well \citep[see also][]{Rattenbury+07}. We find that, for metal-rich stars, there is a strong gradient of the velocity dispersion with latitude, and on the minor axis at low latitudes it becomes higher than (or comparable to) the one of metal-poor stars \citep[consistently with the findings of][in the inner bulge]{Zoccali17}. We also observe an asymmetry with respect to $\ell=0$ for metal-rich stars, with higher values of the dispersion at positive Galactic longitudes. Fig. \ref{fig:sigmab_feh} shows instead the velocity dispersion along Galactic latitude, $\sigma_{\mu_b}$. Once again, we note that metal-poor stars are kinematically hotter, and they do not exhibit a clear coherent pattern over the sky.

To better illustrate the different kinematics of stars with different chemistry, in the upper panel of Fig. \ref{fig:sigmal_sigmab_feh} we plot the velocity dispersion along Galactic longitude (left panel) and Galactic latitude (right panel) as a function of metallicity, for different slices of Galactic latitude $b$. In both cases, we clearly see that metal-rich stars are kinematically colder \citep[see also][]{Arentsen20}. For a given metallicity, stars closer to the Galactic plane show lower values of velocity dispersions (except for the most metal-poor star sample, which also has the largest uncertainties). Finally, we find that the difference in velocity dispersion between the stars at low Galactic latitude and high Galactic latitude increases with metallicity. Following the approach outlined in Section \ref{sec:rot_curves}, we can estimate the dependence of the proper motion dispersions on the ages of the stars, as shown in the lower panel of Fig. \ref{fig:sigmal_sigmab_feh}. Younger stars are kinematically colder, and the highest values of the dispersions ($\sigma_{mu_{\ell*}}\sim 3$ mas yr$^{-1}$, $\sigma_{\mu_b}\sim 2.75$ mas yr$^{-1}$) are obtained for stars older than 13 Gyr.  

The results shown in Figures \ref{fig:sigmal_feh}-\ref{fig:sigmal_sigmab_feh} closely mirror the metallicity dependent radial velocity dispersion variations noted by \citet{Wylie21} in their Figure 26.  Those authors found that stars with [Fe/H] $>$ $-$0.5 had strongly peaked radial velocity dispersions near $\ell$ = 0, but only for fields with 3 $<$ $|$b$|$ $<$ 6.  Additionally, \citet{Wylie21} showed that the difference in radial velocity dispersion between low and high latitude fields increases strongly with increasing [Fe/H].  Therefore, we confirm their result that the more metal-poor stars have flatter velocity dispersion profiles as a function of latitude.

In Fig. \ref{fig:dispratio_feh} we plot the dispersion ratio as a function of metallicity and sky position, defined as the ratio between the velocity dispersion along longitude and along latitude.  Similar to Figure 6 of \citet{Sanders19}, we find a significantly lower dispersion ratio for sight lines near 0$^\circ$ $<$ $\ell$ $<$ $+$5$^\circ$ and $b$ $<$ -6$^\circ$ compared to adjacent fields.  This feature appears to be related to the influence of the X-shape structure in the bulge and is only prominent at [Fe/H] $>$ 0.

In Fig. \ref{fig:correlation_feh}, we plot the correlation coefficient between Galactic proper motions (Eq. \ref{eq:pm_corr}), for the different spatial and metallicity bins. The dipole pattern (which becomes a quadrupole when having access also to the northern Galactic bulge) is a sign of radial alignment towards the Galactic Centre, with a stronger amplitude of the correlation at $\ell > 0$ due to the orientation of the bar \citep[see][]{Clarke19, Sanders19}. By looking at the different metallicity bins, we see how, similarly to Fig. \ref{fig:angular_v_2d_feh} and Fig. \ref{fig:sigmal_feh}, the dipole patterns start to become apparent for {\Feh} $\gtrsim-0.7$ dex, with an amplitude $|\rho_{\ell b}| \sim 0.1$. The maximum amplitude of $|\rho_{\ell b}| \sim 0.2$ is attained for {\Feh} $\gtrsim 0.1$ dex. Interestingly, the asymmetry due to geometric effects at positive Galactic longitudes is most evident for the {\Feh} $=0.2$ bin, but the absence of BDBS fields at $\ell > 3^\circ$, $b<-8^\circ$ hampers the possibility to study this in more details. 

\begin{figure}[]
\centering
\includegraphics[width=0.7\columnwidth]{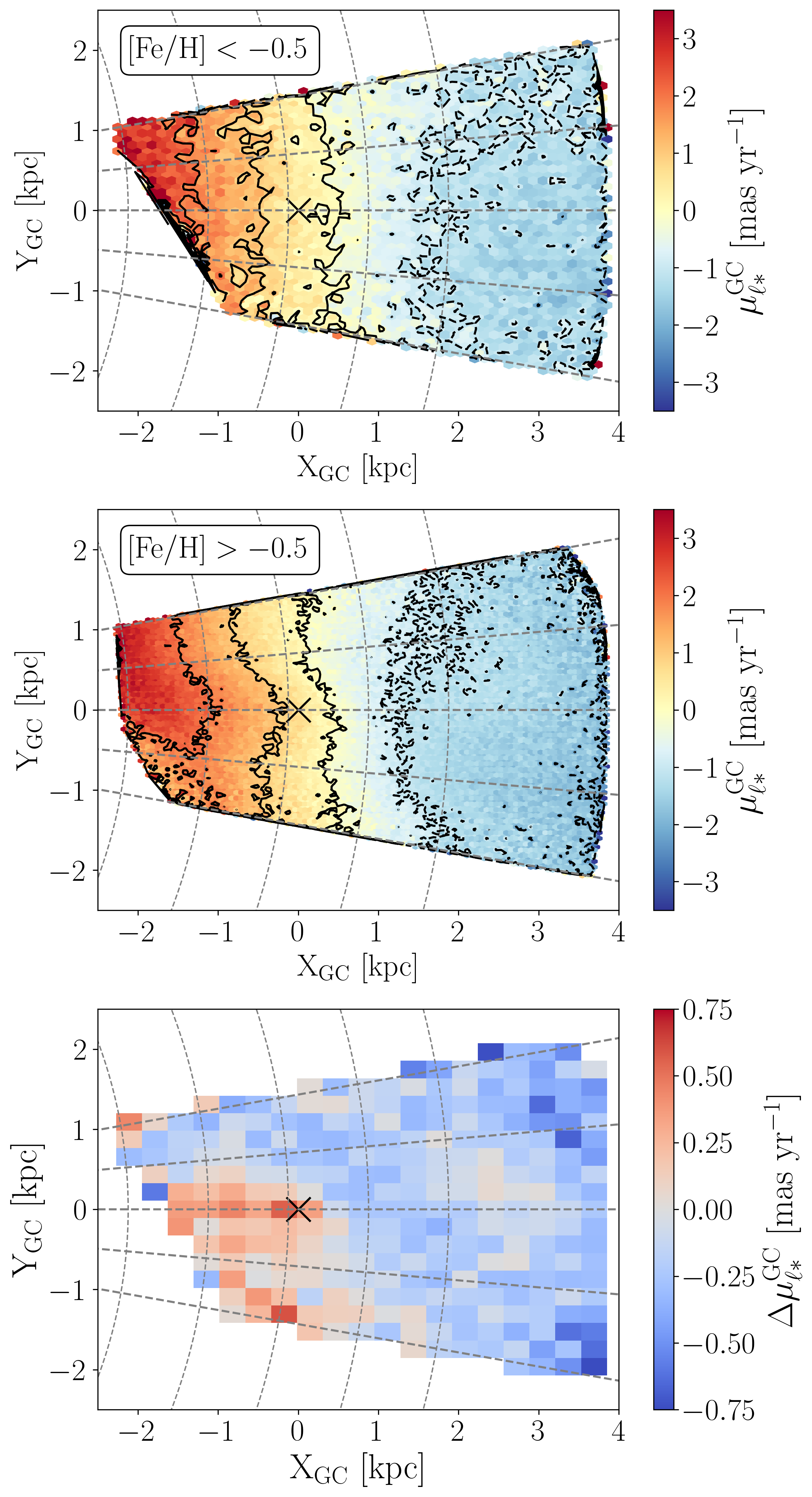}

\caption{Top-down view of the sample of RC stars with {\Feh} $< -0.5$ dex (top panel, $\sim3\times 10^5$ stars) and {\Feh} $> -0.5$ dex (middle panel, $\sim 2.3\times 10^6$ stars), color-coded by the proper motion along Galactic longitude corrected by the motion of the Sun, $\mu_{\ell*}^\mathrm{GC}$. Black lines correspond to lines of constant $\mu_{\ell*}^\mathrm{GC} = (-2, -1, 0, 1, 2)$ mas yr$^{-1}$. The bottom panel is color-coded by the difference in the mean proper motion for metal-rich ({\Feh} $> -0.5$ dex) and metal-poor stars ({\Feh} $< -0.5$ dex) in each spatial bin. The black cross denotes the position of the Galactic Centre, and grey dashed lines are the same as in Fig. \ref{fig:RC_xy_vlvb}.}
\label{fig:XY_feh_pml}
\end{figure}

\begin{figure}[]
\centering
\includegraphics[width=0.7\columnwidth]{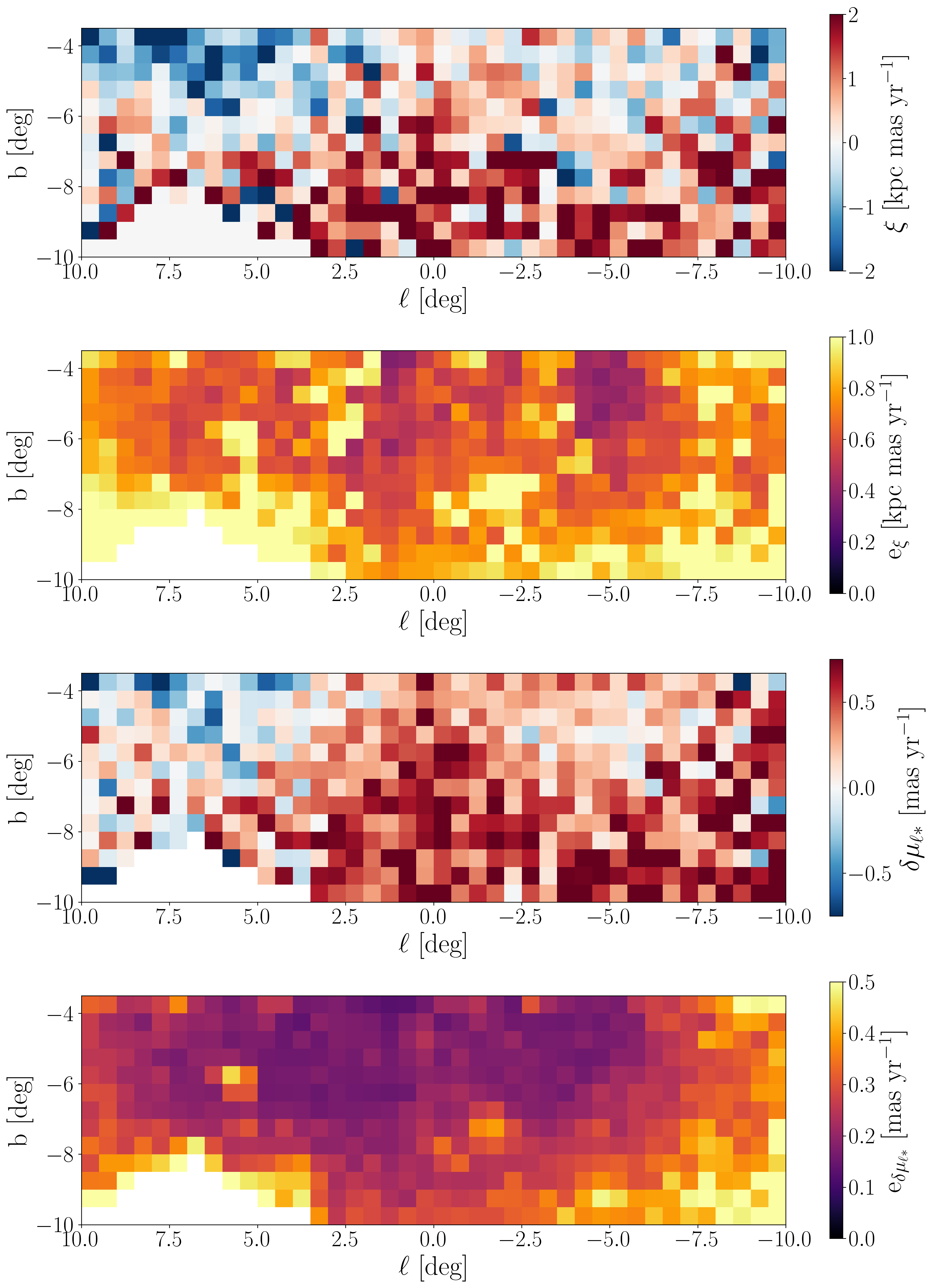}

\caption{First panel: separation amplitude $\xi$. Second panel: error on the separation amplitude. Third panel: Mean longitudinal proper motion difference between metal-rich and metal-poor stars at a distance of $\sim 8$ kpc, $\delta\mu_{\ell*}$. Fourth panel: error on the $\delta\mu_{\ell*}$ metric. All the plots are shown as a function of Galactic coordinates, and the angular bins have sizes $0.5^\circ\times0.5^\circ$.}
\label{fig:xi_2d}
\end{figure}

\subsection{Kinematic fractionation}
\label{sec:kin_fract}

In the previous sections, we noted that the most striking transition features due to the presence of the bar are most evident for [Fe/H] $\gtrsim -0.5$ dex. In this section, we, therefore, decide to split the sample of RC stars into two large metallicity bins, defining metal-poor stars as RC stars with [Fe/H] $<-0.5$ dex, and metal-rich RC stars as those with [Fe/H] $>-0.5$ dex. We will now make a qualitative comparison between our results and the N-body + smoothed-hydrodynamics star-forming simulations of \citet{Debattista17}, following the approach outlined in \citet{Gough-Kelly22}. We note that the metrics introduced in \citet{Gough-Kelly22} to investigate kinematic fractionation through proper motions in the bulge are defined not in terms of metallicity but of ages, for young ($< 7 $ Gyr) and old ($> 9$ Gyr) stars, as ages are more natural units than [Fe/H] for their simulations, due to the lack of chemical mixing.

To further investigate the impact of the bar on the kinematics of stars with different metallicities, in Fig. \ref{fig:XY_feh_pml} we plot the mean proper motion along longitude $\mu_{\ell *}^{GC}$ (corrected for the motion of the Sun) as a function of Galactocentric Cartesian coordinates $(X_\mathrm{GC},Y_\mathrm{GC})$ for metal-poor ({\Feh} $<-0.5$ dex, top panel) and metal-rich ({\Feh} $>-0.5$ dex, middle panel) stars. Confirming the predictions of \citet{Gough-Kelly22} when inspecting the kinematics of young and old stars in N-body+smoothed particle hydrodynamic simulations (see their figure 5), we find that the asymmetry in the velocity field around $\pm Y_\mathrm{GC}$ is clearly observed in the metal-rich sample, while metal-poor stars show an axisymmetric rotation field. The non-axisymmetric pattern observed for more metal-rich stars is a clear signature of the presence of the bar, as shown by the stronger longitudinal variation of the proper motions compared to the metal-poor stars.

To gain more insight into the kinematic differences between metal-rich and metal-poor stars, in the bottom panel of Fig. \ref{fig:XY_feh_pml} we plot, in each bin of Galactocentric Cartesian coordinates, the quantity $\Delta\mu_{\ell*}^\mathrm{GC}$:
\begin{equation}
    \label{eq:deltamul}
    \Delta\mu_{\ell*}^\mathrm{GC} \equiv \langle \mu_{\ell*}^\mathrm{GC, MR} \rangle - \langle \mu_{\ell*}^\mathrm{GC, MP} \rangle
\end{equation}
where $\langle \mu_{\ell*}^\mathrm{GC, MR}\rangle$ and $\langle \mu_{\ell*}^\mathrm{GC, MP}\rangle$ are, respectively, the mean value of the proper motions along Galactic longitude for metal-rich and metal-poor stars in each spatial bin.
We observe a good qualitative agreement with the simulations from \citet{Gough-Kelly22}: the near positive peak in $\Delta\mu_{\ell*}^\mathrm{GC}$ is at $(X_\mathrm{GC}, Y_\mathrm{GC}) \sim (-1, 0)$ kpc with an extended tail towards negative values of Galactic longitude, while the negative peak around $(Y_\mathrm{GC}) \sim 1$ kpc extends to positive longitudes. We stress here that this is only a qualitative comparison between the observed and predicted trends, and we postpone to further work a more rigorous and careful interpretation of our data in light of realistic models of the Galactic bulge. The simple adopted definitions of metal-poor and metal-rich stars might result in overlapping age distributions, which might cause the differences in the spatial distribution between our data and the clear dipole pattern shown in \citet{Gough-Kelly22}. Finally, we note that Fig. \ref{fig:XY_feh_pml} does not change significantly if we consider vertical slices in $Z_\mathrm{GC}$.

Following \citet{Gough-Kelly22}, we define the separation amplitude $\xi$ as the integral along the line of sight of the difference in mean proper motion between the metal-rich and metal-poor stars:
\begin{equation}
    \label{eq:xi}
    \xi = \delta d \cdot \sum_{d = d_1}^{d_2} \Delta \mu_{\ell *}^\mathrm{GC}(d) \ .
\end{equation}
To facilitate the comparison with \citet{Gough-Kelly22}, we fix the parameters $\delta d = 0.5$ kpc, $d_1 = 6$ kpc, $d_2 = 10$ kpc, and we bin in distance using a fixed bin size of $0.5$ kpc.  In the top panel of Fig. \ref{fig:xi_2d}, we plot the separation amplitude $\xi$ as a function of Galactic coordinates, across the BDBS footprint. The observed projected pattern matches qualitatively the one inferred from simulations \citep[see the middle panel in figure 4 of][]{Gough-Kelly22}: the separation amplitude $\xi$ is negative for $\ell \gtrsim 2.5^\circ$, and positive for the other line of sights. The variations of $\xi$ across the Galactic bulge are due to the difference in the intrinsic velocity distributions of the metal-poor and metal-rich populations, tracing different kinematic structures. The presence of negative values of $\xi$ in our data is likely to be due to the absence of metal-poor stars in the closest distance bins (see Fig. \ref{fig:XY_feh_pml}), due to the longitudinal dependence of the colour cuts implemented in \citet{Johnson22} to minimize foreground contamination from the stellar disk.

We compute the error on the separation amplitude, $e_\xi$, using equation 3 in \citet{Gough-Kelly22}. The second panel of Fig. \ref{fig:xi_2d} shows the separation amplitude as a function of Galactic coordinates. The values of $e_\xi$ are driven by the number of stars in each bin, and we find that the overall distribution loosely traces the underlying one for the metal-rich stars.

To quantify the effect of the kinematic fractionation in the southern Galactic bulge, we also define the metric $\delta \mu_{\ell*}$, following \citet{Gough-Kelly22}:
\begin{equation}
    \label{eq:dmul8}
    \delta\mu_{\ell*} = \langle\mu_{\ell*}^\mathrm{GC,MR}\rangle_{8\mathrm{kpc}} - \langle\mu_{\ell*}^\mathrm{GC,MP}\rangle_{8\mathrm{kpc}} \ ,
\end{equation}
where the mean of the proper motions is computed on a large distance bin along the line of sight, from $7$ kpc to $9$ kpc. In the third panel of Fig. \ref{fig:xi_2d} we plot $\delta\mu_{\ell*}$ as a function of Galactic coordinates. The large values of $\delta\mu_{\ell*}$ result from forbidden velocities in the rotation curves (negative longitudinal proper motions at $\ell>0$, and positive values for $\ell < 0$) and are deeply connected to the distribution of radial velocities of stars in the bar \citep{Gough-Kelly22}. A purely axisymmetric distribution of velocities would not produce forbidden velocities, therefore their non-zero observed values are a strong signature of the presence of the bar \citep{Gough-Kelly22}. The observed pattern in the third panel of Fig. \ref{fig:xi_2d} matches again qualitatively the results from \citet{Gough-Kelly22}. The uncertainty on $\delta\mu_{\ell*}$, shown in the fourth panel of Fig. \ref{fig:xi_2d}, follows $e_\xi$, and it is again driven by the underlying number density of {\Feh} $>-0.5$ dex RC stars in our sample.

\section{Discussion and Conclusions}
\label{sec:discuss_concl}

In this work, we have combined {\Gaia} DR3 proper motions with BDBS photometric distances and metallicities, to investigate the chemo-kinematics of a sample of 2.6 million RC stars in the southern Galactic bulge. Our main results can be summarised as follows:

\begin{itemize}

    \item We find that the angular velocity, defined as the slope of the longitudinal proper motion $\mu_{\ell*}$ curve as a function of distance, is highest at $\omega \sim 39$ km s$^{-1}$ kpc$^{-1}$ for stars with [Fe/H] $\sim -0.25$ dex (see Fig.\ref{fig:angular_v_feh}). Metal-poor RC stars exhibit the lowest rotation values ($\omega \sim 29$ km s$^{-1}$ kpc$^{-1}$). Surprisingly, the angular velocity is not a monotonic function of metallicity, but it decreases for RC stars with [Fe/H] $\gtrsim -0.25$ dex, and it reaches $\omega \sim 35$ km s$^{-1}$ kpc$^{-1}$ for [Fe/H] $\sim 0.5$ dex. When integrated over all metallicities, the angular velocity peaks on the minor axis closer to the plane (Fig. \ref{fig:angular_v_2d}). When plotted as a function of both metallicity and sky position (Fig. \ref{fig:angular_v_2d_feh}), the angular velocity peaks at low latitudes for stars with [Fe/H] $\gtrsim -0.5$ dex, with a clear asymmetry with Galactic longitude, possibly due to the orientation of the near side of the bar. Stars with lower metallicities do not exhibit a coherent angular velocity pattern on the sky.
    
    \item Proper motion dispersions along Galactic longitude and latitude are maximum towards the Galactic Centre for stars with [Fe/H] $\gtrsim -0.5$ dex, because of the Galactic potential well (Fig. \ref{fig:sigmal_feh} and \ref{fig:sigmab_feh}, respectively). Metal-poor stars are kinematically hotter, with proper motion dispersions $\sim 3$ mas yr$^{-1}$, while the mean dispersion for metal-rich stars is $\sim 2$ mas yr$^{-1}$. If we plot the dispersions as a function of metallicity for different latitude bins (Fig. \ref{fig:sigmal_sigmab_feh}) we see that, for a given metallicity bin, RC stars are kinematically hotter closer to the plane. Also, the spread in dispersion between the stars at low latitudes and high latitudes increases with increasing metallicity. 
    
    \item The correlation between Galactic proper motions clearly shows a quadrupole pattern corresponding to radial alignment for stars with [Fe/H] $\gtrsim -0.5$ dex (Fig. \ref{fig:correlation_feh}). The prominence of the correlation amplitude at $\ell >0^\circ$ is a consequence of the orientation of the bar.

    \item Based on the striking difference in the sky plots, we split our sample of RC stars into metal-poor ([Fe/H] $<-0.5$ dex) and metal-rich ([Fe/H] $>-0.5$ dex) stars. We compare our results to the simulations presented in \citet{Gough-Kelly22}, which analyze the predicted proper motion trends for young and old stars in the Galactic bulge. We observe a non-axisymmetric pattern around the Galactic center in the proper motions along Galactic longitude only for the metal-rich sample, which is a clear sign of the bar. More generally, we find that the footprint of the bar is clearly present in the metal-rich sample, and totally absent in the metal-poor one.

    \end{itemize}

Our results clearly demonstrate that RC stars with different metallicities have dramatically different kinematics. In particular, metal-rich stars are kinematically colder, and their orbits show a clear imprint of the Galactic bar. On the other hand, metal-poor RC stars have larger velocity dispersions, and exhibit random motions over the bulge, without showing any radial alignment towards the Galactic Centre, or clear asymmetries with Galactic longitude that could be due to the bar's orientation. At the moment, we are not able to assess whether the more metal-poor stars trace a weaker bar, or if they constitute the so-called classical bulge, assembled over cosmic time by galaxy mergers. A more detailed comparison with cosmological simulations is needed to assess the contribution of the different formation scenarios to the observed kinematics of the stellar populations in the Galactic bulge, and we postpone this to a future dedicated study.

Our results also help us interpret age--metallicity relations in the Galactic bulge.
Though others have proposed a flat age--metallicity relation in the bulge (e.g., \citealt{Bensby17}), the present chemodynamical results support membership of the highest-metallicity star to the bar rather than bulge. This hypothesis is more broadly consistent with the results of \citet{Joyce+23}, who found that, though a handful of the highest--metallicity, micro-lensed subdwarfs identified as belonging to the bulge region may be as young as 2 Gyr, the majority are not. These results, in combination with the kinematic--metallicity correlations presented here suggest that the most metal--rich stars belong to the bar rather than the bulge.
Further, based on Figure \ref{fig:correlation_feh}, the bar orbits become visible for stars with [Fe/H] $> -0.2$. According to Figure \ref{fig:angular_v_feh}, stars in this metallicity range have ages less than 12 Gyr, with the youngest stars having ages around 8.4 Gyr. We note, however, that the bar structure itself may be younger than the stars comprising it.

The values of the angular velocity we determine for RC stars are in agreement with findings for RRL stars \citep{Du+20}, but considerably slower than what was previously found for younger bulge stars \citep[e.g.][]{Sanders19, Du+20}.
In the upcoming years, new data releases by {\Gaia} will provide more precise proper motions thanks to the improved observational baseline. In addition to that, a longer baseline will also improve the accuracy of the astrometric solution, especially in crowded fields, allowing a better call of proper motion systematics. The Vera C. Rubin Telescope will also monitor the Milky Way bulge, and derive proper motions down to the Main Sequence Turnoff \citep{Gonzalez18}. In addition, large-scale multifiber spectroscopic facilities such as MOONS \citep{MOONS} and 4MOST \citep{4MOST} will measure radial velocities for millions of giants in the bulge, allowing a full three-dimensional kinematic analysis when combined with photometric and spectroscopic metallicities.

\vspace{5mm}
\begin{acknowledgements}

T.M. thanks F. Fragkoudi for interesting discussions.

M.J. gratefully acknowledges funding of MATISSE: \textit{Measuring Ages Through Isochrones, Seismology, and Stellar Evolution}, awarded through the European Commission's Widening Fellowship.  
This project has received funding from the European Union's Horizon 2020 research and innovation programme.

Data used in this paper comes from the Blanco DECam Survey Collaboration. This project used data obtained with the Dark Energy Camera (DECam), which was constructed by the Dark Energy Survey (DES) collaboration. Funding for the DES Projects has been provided by the U.S. Department of Energy, the U.S. National Science Foundation, the Ministry of Science and Education of Spain, the Science and Technology Facilities Council of the United Kingdom, the Higher Education Funding Council for England, the National Center for Supercomputing Applications at the University of Illinois at Urbana-Champaign, the Kavli Institute of Cosmological Physics at the University of Chicago, the Center for Cosmology and Astro-Particle Physics at the Ohio State University, the Mitchell Institute for Fundamental Physics and Astronomy at Texas A\&M University, Financiadora de Estudos e Projetos, Funda\c{c}\~{o} Carlos Chagas Filho de Amparo \'a Pesquisa do Estado do Rio de Janeiro, Conselho Nacional de Desenvolvimento Cient\'{\i}fico e Tecnol\'ogico and the Minist\'erio da Ci\^encia, Tecnologia e Inovac\~{a}o, the Deutsche Forschungsgemeinschaft, and the Collaborating Institutions in the Dark Energy Survey. The Collaborating Institutions are Argonne National Laboratory, the University of California at Santa Cruz, the University of Cambridge, Centro de Investigaciones En\'ergeticas, Medioambientales y Tecnol\'ogicas-Madrid, the University of Chicago, University College London, the DES-Brazil Consortium, the University of Edinburgh, the Eidgen\"ossische Technische Hochschule (ETH) Z\"urich, Fermi National Accelerator Laboratory, the University of Illinois at Urbana-Champaign, the Institut de Ci\'oncies de l'Espai (IEEC/CSIC), the Institut de F\'{\i}sica d’Altes Energies, Lawrence Berkeley National Laboratory, the Ludwig-Maximilians Universit\"at M\"unchen and the associated Excellence Cluster Universe, the University of Michigan, the National Optical Astronomy Observatory, the University of Nottingham, the Ohio State University, the OzDES Membership Consortium the University of Pennsylvania, the University of Portsmouth, SLAC National Accelerator Laboratory, Stanford University, the University of Sussex, and Texas A\&M University. Based on observations at Cerro Tololo Inter-American Observatory (2013A-0529; 2014A-0480; PI: Rich), National Optical Astronomy Observatory, which is operated by the Association of Universities for Research in Astronomy (AURA) under a cooperative agreement with the National Science Foundation. 

This work has made use of data from the European Space Agency (ESA) mission
{\it Gaia} (\url{https://www.cosmos.esa.int/gaia}), processed by the {\it Gaia}
Data Processing and Analysis Consortium (DPAC,
\url{https://www.cosmos.esa.int/web/gaia/dpac/consortium}). Funding for the DPAC
has been provided by national institutions, in particular the institutions
participating in the {\it Gaia} Multilateral Agreement.

\emph{Software used:} numpy \citep{numpy},
          matplotlib \citep{matplotlib},
          scipy \citep{scipy},
          astropy \citep{astropy13, astropy18},
          TOPCAT \citep{topcat1, topcat2}

\end{acknowledgements}

\bibliography{main}{}
\bibliographystyle{aa}




%
%
\end{document}